\documentclass[apj,iop]{emulateapj}
\usepackage[colorlinks,citecolor=blue,linkcolor=blue,urlcolor=blue]{hyperref}

\newcommand {\be} {\begin{equation}}
\newcommand {\ee} {\end{equation}}
\newcommand {\bea} {\begin{eqnarray}}
\newcommand {\eea} {\end{eqnarray}}

\begin{document}

\title{On the distribution of particle acceleration sites in plasmoid-dominated relativistic magnetic reconnection}
\shorttitle{Distribution of acceleration sites in reconnection}

\author{Krzysztof~Nalewajko\altaffilmark{1,2,3}, Dmitri~A.~Uzdensky\altaffilmark{4}, Beno{\^i}t~Cerutti\altaffilmark{5,6}, Gregory~R.~Werner\altaffilmark{4}, Mitchell~C.~Begelman\altaffilmark{2,7}}
\shortauthors{Nalewajko et~al.}

\altaffiltext{1}{Kavli Institute for Particle Astrophysics and Cosmology, SLAC National Accelerator Laboratory, Stanford University, 2575 Sand Hill Road M/S 29, Menlo Park, CA 94025, USA; {\tt knalew@stanford.edu}}
\altaffiltext{2}{JILA, University of Colorado and National Institute of Standards and Technology, 440 UCB, Boulder, CO 80309, USA}
\altaffiltext{3}{NASA Einstein Postdoctoral Fellow}
\altaffiltext{4}{Center for Integrated Plasma Studies, Physics Department, University of Colorado, UCB 390, Boulder, CO 80309-0390, USA}
\altaffiltext{5}{Department of Astrophysical Sciences, Princeton University, Princeton, NJ 08544, USA}
\altaffiltext{6}{Lyman Spitzer Jr. Fellow}
\altaffiltext{7}{Department of Astrophysical and Planetary Sciences, University of Colorado, UCB 391, Boulder, CO 80309-0391, USA}

\begin{abstract}
We investigate the distribution of particle acceleration sites, independently of the actual acceleration mechanism, during plasmoid-dominated, relativistic collisionless magnetic reconnection by analyzing the results of a particle-in-cell numerical simulation.
The simulation is initiated with Harris-type current layers in pair plasma with no guide magnetic field,
negligible radiative losses, no initial perturbation, and using periodic boundary conditions.
We find that the plasmoids develop a robust internal structure, with colder dense cores and hotter outer shells, that is recovered after each plasmoid merger on a dynamical time scale.
We use spacetime diagrams of the reconnection layers to probe the evolution of plasmoids, and in this context we investigate the individual particle histories for a representative sample of energetic electrons.
We distinguish three classes of particle acceleration sites associated with (1) magnetic X-points, (2) regions between merging plasmoids, and (3) the trailing edges of accelerating plasmoids.
We evaluate the contribution of each class of acceleration sites to the final energy distribution of energetic electrons --- magnetic X-points dominate at moderate energies,
and the regions between merging plasmoids dominate at higher energies.
We also identify the dominant acceleration scenarios, in order of decreasing importance --- (1) single acceleration between merging plasmoids, (2) single acceleration at a magnetic X-point, and (3) acceleration at a magnetic X-point followed by acceleration in a plasmoid.
Particle acceleration is absent only in the vicinity of stationary plasmoids.
The effect of magnetic mirrors due to plasmoid contraction does not appear to be significant in relativistic reconnection.
\end{abstract}


\section{Introduction}

Rapid bursts of non-thermal radiation are common phenomena in high-energy astrophysics \citep{2015MNRAS.446.3687P}. Examples include solar flares \citep{2003NewAR..47...53L}, gamma-ray bursts \citep{2004RvMP...76.1143P,2015PhR...561....1K}, flares of blazars \citep{2010ApJ...722..520A,2013MNRAS.430.1324N}, giant flares of magnetars \citep{2008A&ARv..15..225M}, and gamma-ray flares of the Crab Nebula \citep{2011Sci...331..739A,2011Sci...331..736T}. Despite widely different contexts, they pose similar theoretical questions: what is the energy source, how is this energy dissipated, how are particles accelerated, and what is the radiation mechanism. We can address these questions by studying the fundamental properties of plausible physical mechanisms, and comparing the results with observations.

Many high-energy astrophysical environments are strongly magnetized, with a magnetization parameter $\sigma = B^2/(4\pi w)> 1$, where $B$ is the magnetic field strength and $w$ is the relativistic enthalpy. In such environments, magnetic reconnection, operating in the relativistic regime, can be the dominant dissipation mechanism \citep{2009ARA&A..47..291Z,2011SSRv..160...45U,2015SSRv..tmp....3K}. Relativistic magnetic reconnection is a very complex and non-linear process, difficult to describe analytically or semi-analytically \citep{1994PhRvL..72..494B,2003ApJ...589..893L,2005MNRAS.358..113L,2009MNRAS.395L..29G,2011MNRAS.413..333N}. A lot of recent progress has been made through numerical studies, in particular by ab-initio particle-in-cell (PIC) plasma simulations for collisionless plasmas \citep{2001ApJ...562L..63Z,2004PhPl...11.1151J,2005ApJ...618L.111Z,2005PhRvL..95i5001Z,2007PhPl...14g2303D,2007PhPl...14e6503B,2007ApJ...670..702Z,2008ApJ...677..530Z,2008ApJ...682.1436L,2009PhRvL.103g5002J,2011ApJ...741...39S,2011PhPl...18e2105L,2012ApJ...750..129B,2012ApJ...754L..33C,2013ApJ...770..147C,2013ApJ...774...41K,2014ApJ...782..104C,2014ApJ...783L..21S,2014A&A...570A.111M,2014PhRvL.113o5005G,2014arXiv1409.8262W,2015ApJ...806..167G,2015PhRvL.114i5002L}.

Significant recent progress was motivated by the problem posed by the Crab Nebula flares --- how to exceed the theoretical limit for the observed energy of the synchrotron radiation, $\mathcal{E}_{\rm max} \simeq 160\;{\rm MeV}(E/B)$, where $E$ is the electric field strength and $B$ is the magnetic field strength, due to the radiation reaction \citep{2011ApJ...737L..40U,2012ApJ...746..148C}. Implementation of the radiation reaction force in the {\tt Zeltron} PIC code allowed for the self-consistent calculation of radiative signatures (due to the synchrotron process) from relativistic reconnection \citep{2013ApJ...770..147C,2014ApJ...782..104C}. But even in the limit of negligible effect of the radiation reaction, radiation produced during relativistic reconnection has very interesting properties: strong energy-dependent anisotropy (`kinetic beaming') and rapid variability observed in certain directions \citep{2012ApJ...754L..33C}.

In a typical numerical experiment on collisionless magnetic reconnection, the initial conditions include a uniform Harris-type current layer (\citealt{1962Harris}; see \citealt{2003ApJ...591..366K} for detailed description of such layers in the relativistic case).
We can consider a configuration in which the current layer is in the $(x,z)$ plane and the reconnecting magnetic field is along the $x$ axis.
Long and thin current layers are prone to two types of instabilities: (1) a tearing mode \citep[e.g.,][]{1977PhFl...20.1341D} producing a series of plasmoids (magnetic O-points) and magnetic X-points along $x$, which subsequently undergo highly dynamic evolution by a series of mergers; and (2) a drift-kink mode \citep{1996JGR...101.4885Z,1996JGR...10127413P} bending the layer along $z$.
Standard 2.5-dimensional simulations (with 2D position space and 3D momentum space) in the $(x,y)$ plane capture only the tearing mode, neglecting the effects of the drift-kink mode which tends to inhibit particle acceleration \citep{2005ApJ...618L.111Z,2007ApJ...670..702Z}.
Although it is possible to stabilize the drift-kink mode by introducing a guide magnetic field (a uniform $z$-component) \citep{2005PhRvL..95i5001Z,2008ApJ...677..530Z}, the results of recent very large three-dimensional simulations suggest that even in the absence of the guide field the negative effect of the drift-kink mode on non-thermal particle acceleration is only temporary and particle acceleration recovers on sufficiently long time scales \citep{2013ApJ...774...41K,2014ApJ...783L..21S,2014PhRvL.113o5005G,2015ApJ...806..167G}.

Several mechanisms of particle acceleration at different acceleration sites in relativistic magnetic reconnection have been identified \citep[e.g.,][]{2012SSRv..173..521H}. The most direct mechanism, that is especially efficient at magnetic X-points, is due to the reconnection electric field $E_z$ induced by reconnecting magnetic field $B_x$ inflowing along $y$ \citep{2001ApJ...562L..63Z,2003ApJ...586...72L,2008ApJ...682.1436L}.
Another mechanism operates between two approaching plasmoids shortly before they merge --- their motion along $x$ induces a strong electric field $E_z$ of opposite sign to the main reconnection electric field (hence called the anti-reconnection electric field; \citealt{2010ApJ...714..915O}, see also \citealt{2004PhPl...11.1151J}).
In addition, a first-order Fermi mechanism may operate within the plasmoids due to their contraction \citep{2006Natur.443..553D}, and more generally due to the curvature drifts \citep{2014PhPl...21i2304D,2014PhRvL.113o5005G,2015ApJ...806..167G}.
Recent results of very large PIC simulations of relativistic magnetic reconnection have demonstrated that power-law energy distributions of particles arise quite efficiently \citep{2014ApJ...783L..21S,2014PhRvL.113o5005G,2014arXiv1409.8262W,2015ApJ...806..167G}.

Since at least three acceleration mechanisms have been identified in previous numerical studies of magnetic reconnection, it is necessary to evaluate their relative importance.
In this work, we attempt to address this issue by investigating a \emph{representative} sample of energetic electrons (i.e., all individually tracked particles that exceed a certain energy threshold).
We propose a simple method of identifying the main acceleration episodes for each particle, and use the distribution of these episodes to make a crude classification of the main acceleration sites.
The results presented here are based on a single 2.5-dimensional PIC simulation of relativistic magnetic reconnection in an electron-positron plasma, initiated from an unperturbed Harris-type current layer, and performed with the {\tt Zeltron} code. In Section \ref{sec_setup}, we present the simulation setup. In Section \ref{sec_res}, we present the detailed results of our simulation, including the analysis of individual histories for a representative sample of high-energy particles. A discussion follows in Section \ref{sec_dis}, and conclusions in Section \ref{sec_con}.

\section{Simulation setup}
\label{sec_setup}

We perform a 2.5-dimensional PIC simulation of relativistic magnetic reconnection in pair plasma using the {\tt Zeltron} code\footnote{\url{http://benoit.cerutti.free.fr/Zeltron/}} \citep{2013ApJ...770..147C} in an $(x,y)$ domain with periodic boundaries. The domain size is $L_x = L_y = 800\rho_0$, where $\rho_0 = m_{\rm e}c^2/(eB_0)$ is the nominal gyroradius and $B_0$ is the characteristic value of the reconnecting magnetic field strength.
The coordinate limits are $0 \le x < L_x$ and $0 \le y < L_y$.
The initial condition is a pair of relativistic Harris-type current sheets \citep{2003ApJ...591..366K} parallel to each other, oriented along the $x$ coordinate (with the current flowing along $\pm z$), and centered at $y = 0.25L_y$ and $y = 0.75L_y$.
We set the initial magnetic field as $B_x(y) = -B_0\tanh[(y-0.25L_y)/\delta]$ for $0 \le y < 0.5L_y$, and $B_x(y) = B_0\tanh[(y-0.75L_y)/\delta]$ for $0.5L_y \le y < L_y$, where $\delta$ is the characteristic half-thickness of the initial current layers.
We do not have any kind of initial perturbation of the magnetic field (hence $B_y = 0$), nor any guide magnetic field (hence $B_z = 0$).
The gradient of magnetic field $B_x$ along $y$ is supported by the plasma pressure (pressure balance) and by the electric current (Amp{\'e}re's law), both provided by a drifting pair plasma component characterized by initial dimensionless temperature $\Theta_{\rm d} = kT_{\rm d}/(m_{\rm e}c^2) = 1$ and by dimensionless drift velocity in the $\pm z$ direction $\beta_{\rm d} = 0.6$.
The number density of the drifting plasma (including both electrons and positrons) is given by $n_{\rm d}(y) = n_{\rm d,0}\cosh^{-2}[(y-0.25L_y)/\delta]$ for $0 \le y < 0.5L_y$, and $n_{\rm d}(y) = n_{\rm d,0}\cosh^{-2}[(y-0.75L_y)/\delta]$ for $0.5L_y \le y < L_y$, where $n_{\rm d,0} = \gamma_{\rm d}B_0^2/(8\pi\Theta_{\rm d}m_{\rm e}c^2)$ is the peak number density and $\gamma_{\rm d} = (1-\beta_{\rm d}^2)^{-1/2} = 1.25$ is the drift Lorentz factor.
The temperature and the drift velocity of the drifting electrons and positrons determine the thickness of the current layer $\delta = 2\Theta_{\rm d}\rho_0/(\gamma_{\rm d}\beta_{\rm d}) = (8/3)\rho_0$ \citep{2003ApJ...591..366K}.
We also have a background plasma component of dimensionless temperature $\Theta_{\rm b} = kT_{\rm b}/(m_{\rm e}c^2) = 1$ and uniform number density $n_{\rm b}$ corresponding to a background magnetization of $\sigma_0 = B_0^2/(4\pi n_{\rm b}m_{\rm e}c^2) = 16$, and hence to a drifting-to-background density ratio of $n_{\rm d,0}/n_{\rm b} = \gamma_{\rm d}\sigma_0/(2\Theta_{\rm d}) = 10$.
We do not include the effect of radiation reaction from the synchrotron process.
The simulation grid has $N_x\times N_y = 2048^2$ cells, so that the cell size is $\Delta x = \Delta y = L_x/N_x \simeq 0.4\rho_0 \simeq 0.15\delta$.
Initially, each cell contains 256 particles including background and drifting populations of both electrons and positrons.
The density structure for drifting particles is obtained by adjusting particle weights; the weights of background particles are uniform.
The simulation is run for $N_t = 10^4$ time steps, with each time step given by $\omega_0\Delta t = 0.9\Delta x/\sqrt{2}\rho_0 \simeq 0.25$, where $\omega_0 = eB_0/(m_{\rm e}c) = c/\rho_0$ is the nominal cyclotron frequency.
Therefore, the total duration of the simulation is $\omega_0t_{\rm max} = N_t\omega_0\Delta t \simeq 2500$, which corresponds to $ct_{\rm max}/L_x \simeq 3.1$ light-crossing times of the simulation domain.

\begin{figure*}[ht]
\centering
\includegraphics[width=0.8\textwidth]{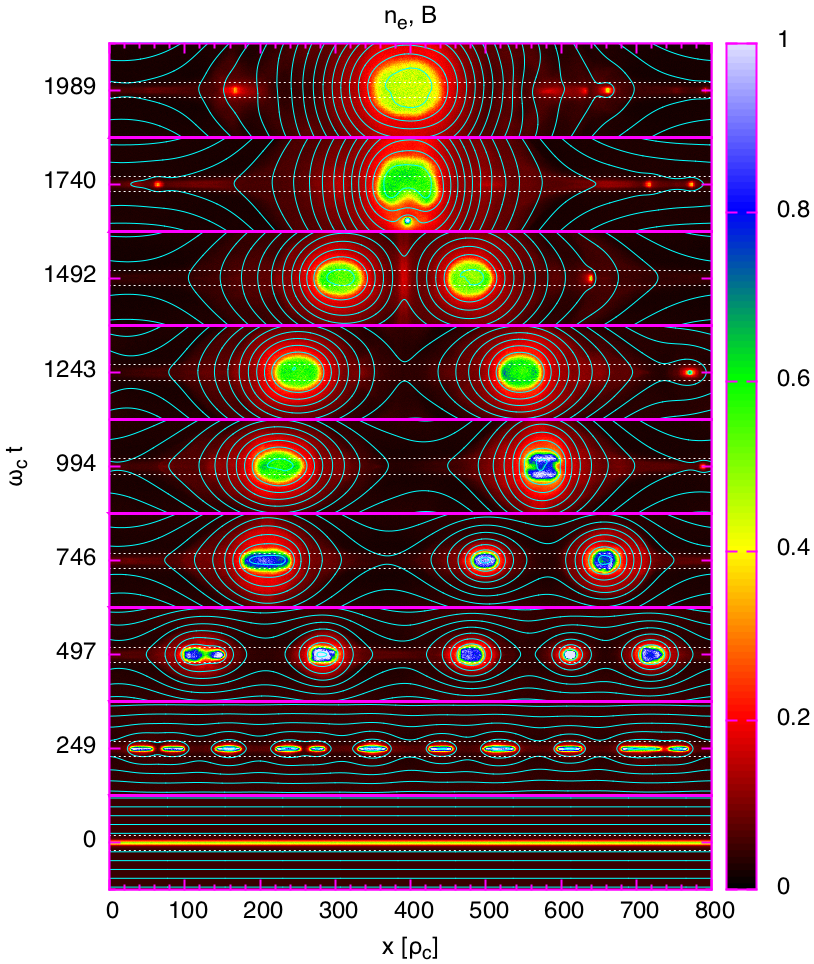}
\caption{Maps in $x,y$ coordinates of the total electron density (\emph{color scale}; arbitrary units) and magnetic field lines (\emph{cyan lines}) taken at regular intervals of simulation time indicated on the left axis. \emph{Dotted white lines} mark the region with $\Delta y = 20\rho_0$, from which the spacetime diagrams were extracted.}
\label{fig_xymap}
\end{figure*}

\section{Results}
\label{sec_res}

Here, we analyze the evolution of just one of the two reconnection layers present in our simulation, the one centered at $y = 0.75L_y$, as the results are generally similar for both layers.
Each layer develops a different set of plasmoids, the details of which are chaotic and unpredictable \emph{a priori} due to the initial random noise in the particle distribution.
In addition to $(x,y)$ snapshots (cuts in $t$), we will rely heavily on $(x,t)$ spacetime diagrams (cuts in $y$ centered on the current sheet). Also, we will focus on the electrons only, as the behavior of the positrons is highly symmetric to that of the electrons (they co-move along $x$ and counter-move along $z$). In Section \ref{sec_res_plasm}, we describe the evolution of plasmoids in the reconnection layer using spacetime diagrams. Then, in Section \ref{sec_res_accel}, we will characterize the acceleration of energetic particles in the context of the plasmoid evolution.

\subsection{Evolution of plasmoids}
\label{sec_res_plasm}

In Figure~\ref{fig_xymap}, we present snapshots of the $(x,y)$ maps of electron number density and magnetic field lines for several simulation times.
Each snapshot shows only a fragment of the simulation domain for $0 \le x \le L_x$ and $550\rho_0 \le y \le 660\rho_0$, centered on the reconnection layer of interest ($y = 500\rho_0$).
As has been demonstrated in numerous similar studies, most of the interesting physical activity is concentrated in the vicinity of the reconnection layers.
At $t = 0$, the system consists of a thin current layer and magnetic field $B_x$, both uniform along $x$.
By $\omega_0t = 249 \simeq 0.31(L_x/\rho_0)$, the tearing-mode instability has disrupted the current layer and produced a chain of plasmoids, also known as magnetic islands or magnetic flux ropes in 3D, characterized by high particle density surrounded by closed magnetic field lines (they are associated with the peaks of the magnetic vector potential $A_z$). They are separated by magnetic X-points characterized by low particle density (the saddle points of $A_z$).
At this early stage, the plasmoids are significantly elongated in the $x$ direction.
Subsequently, the plasmoids are accelerated along $\pm x$, merge in a hierarchical order, and also circularize and maintain a roughly round shape after each merger.
By $\omega_0t = 994 \simeq 1.24(L_x/\rho_0)$ only 2 plasmoids remain of the original set, although small secondary plasmoids emerge occasionally from the major X-point (i.e., the X-point that survives to the end of the simulation).
Because the $x = 0 = L_x$ boundary is periodic, we have the freedom of translating the simulation domain along $x$, and we chose a translation (fixed for the entire simulation time) that sets the major X-point at the boundary.
The location of the major X-point cannot be known \emph{a priori}, as it depends on the entire evolutionary history of the plasmoids.
By choosing such a translation, we can ensure that no plasmoid from the original chain ever crosses the boundary.
At $\omega_0t = 1492 \simeq 1.87(L_x/\rho_0)$, a secondary vertical current layer (oriented along the $y$-direction) becomes visible between the two large plasmoids in an early stage of their merger.
At $\omega_0t = 1740 \simeq 2.18(L_x/\rho_0)$, the final merger is almost complete with a newly formed plasmoid escaping along the $-y$ axis.
Plasmoids forming due to the tearing mode instability of the secondary vertical current sheets were previously seen by \cite{2007PhPl...14g2303D} and \cite{2014ApJ...783L..21S}.
The evolution of plasmoids is balanced in the sense that at any given instance the existing plasmoids are similar to each other and roughly equidistant.\footnote{The evolution of plasmoids is very sensitive to the introduction of initial perturbations. We checked for the effect of long-wavelength initial perturbation of the magnetic field. Even at very modest perturbation amplitude of $0.03 B_0$, we observed the rapid creation of a major plasmoid resulting in a markedly unbalanced subsequent evolution.}

\begin{figure*}[ht]
\centering
\includegraphics[width=0.47\textwidth]{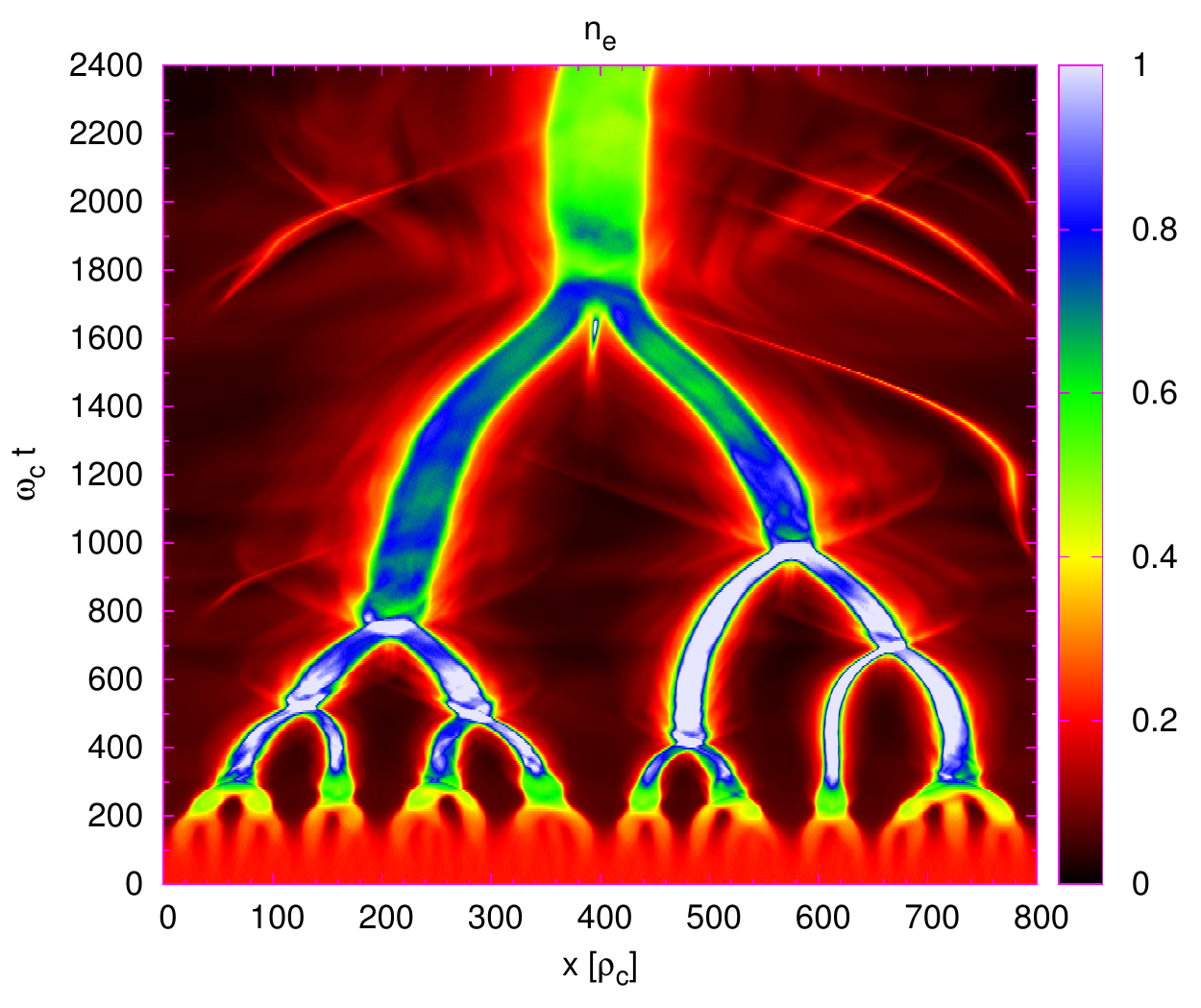}
\includegraphics[width=0.47\textwidth]{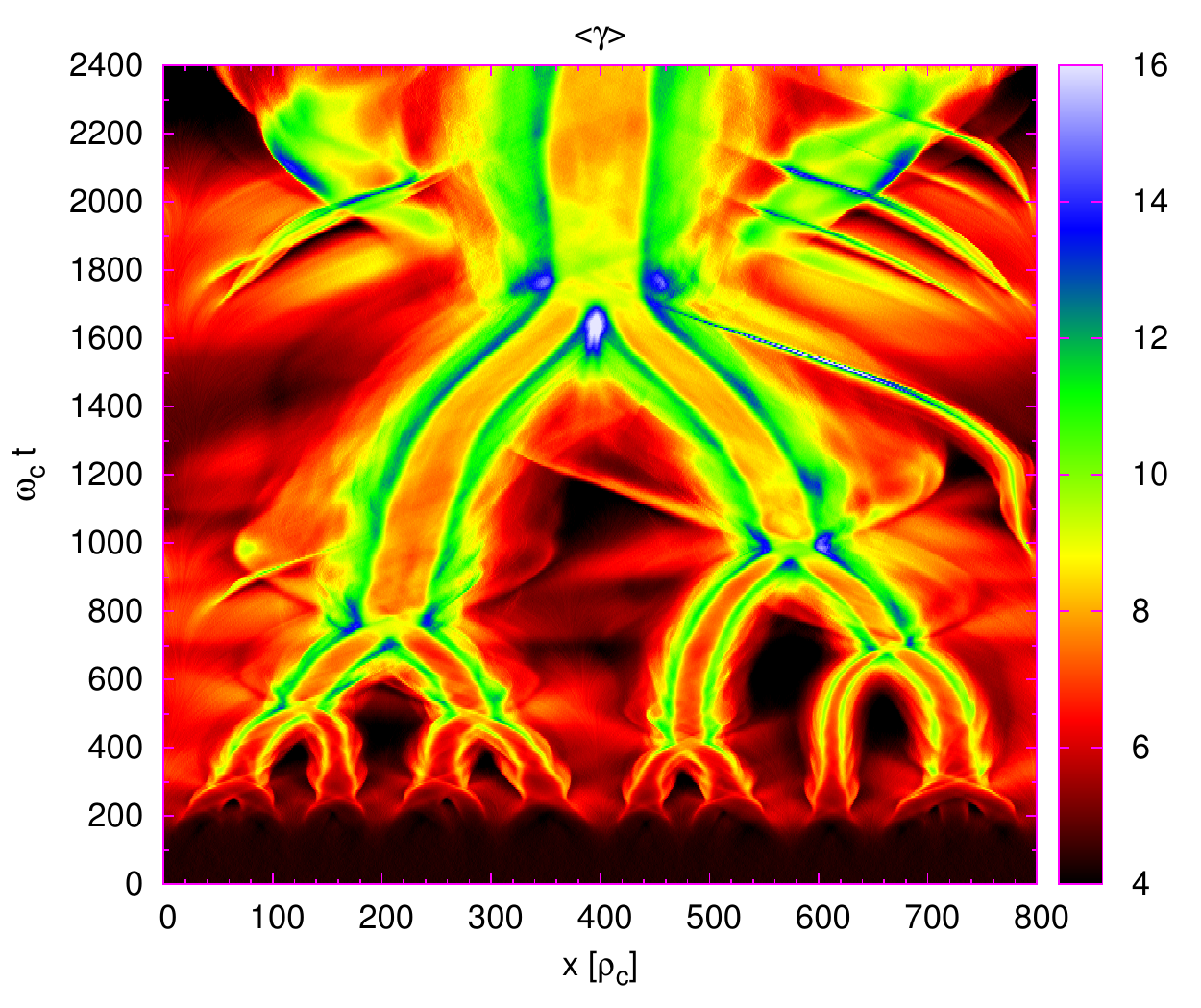}
\includegraphics[width=0.47\textwidth]{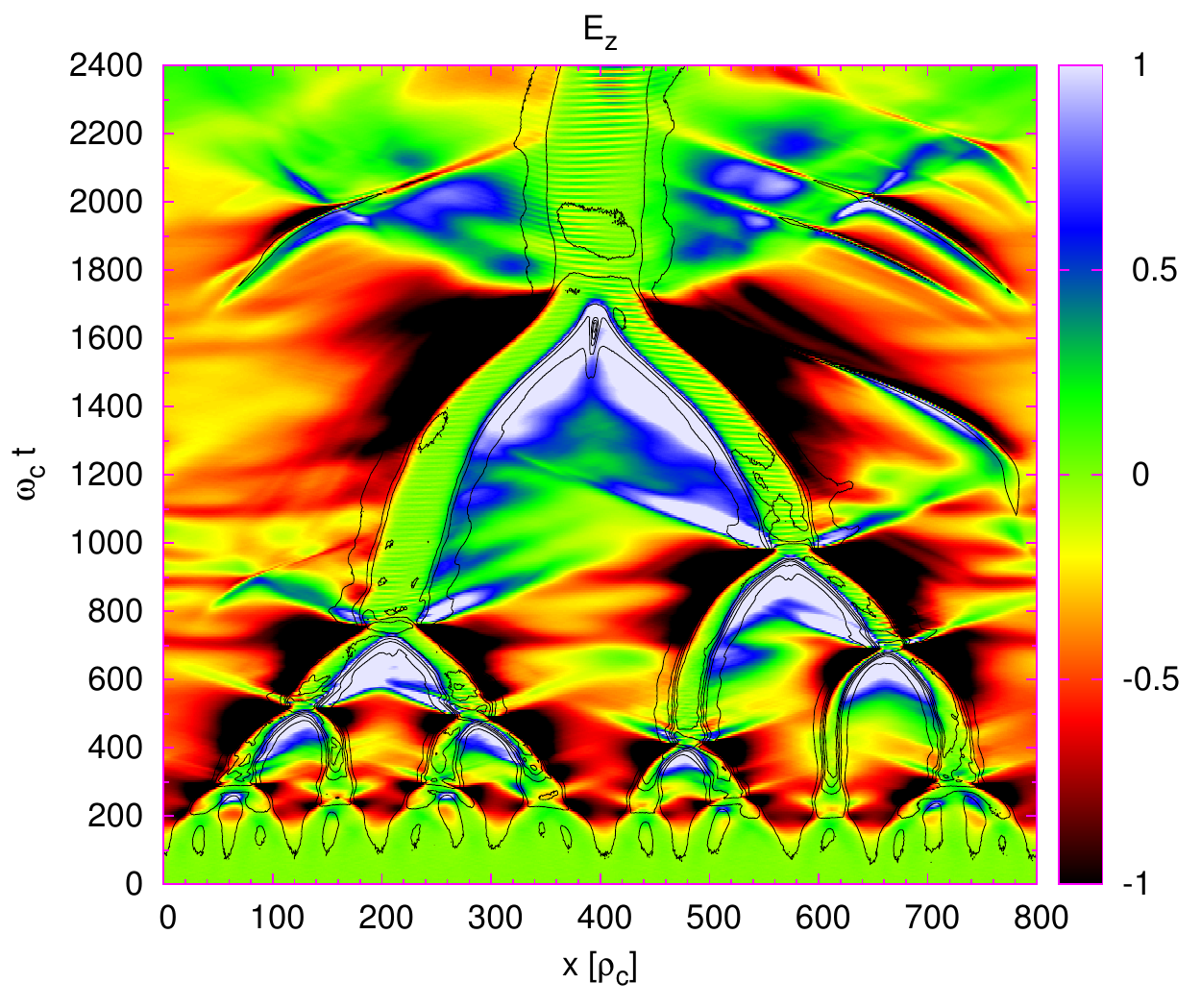}
\includegraphics[width=0.47\textwidth]{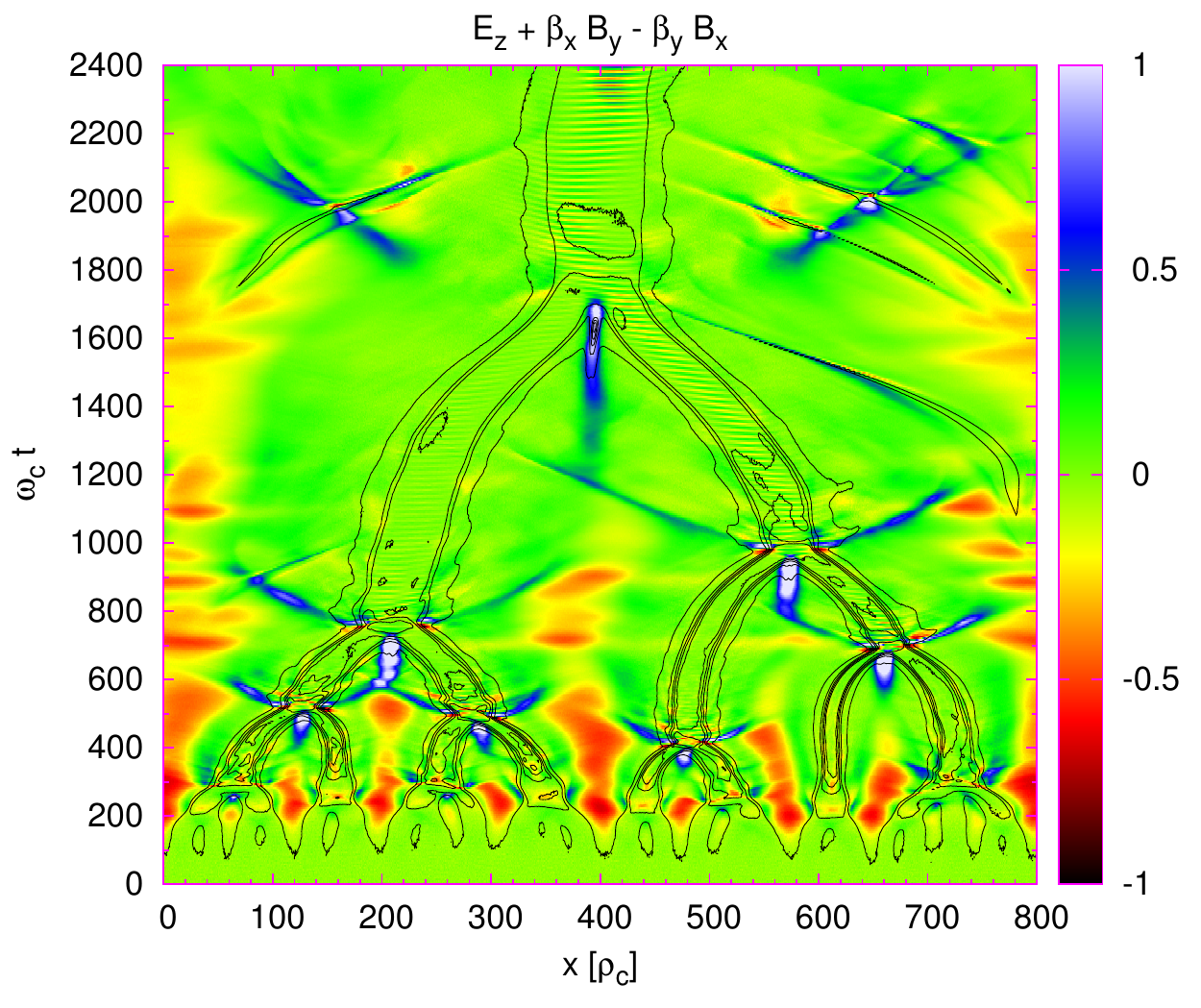}
\includegraphics[width=0.47\textwidth]{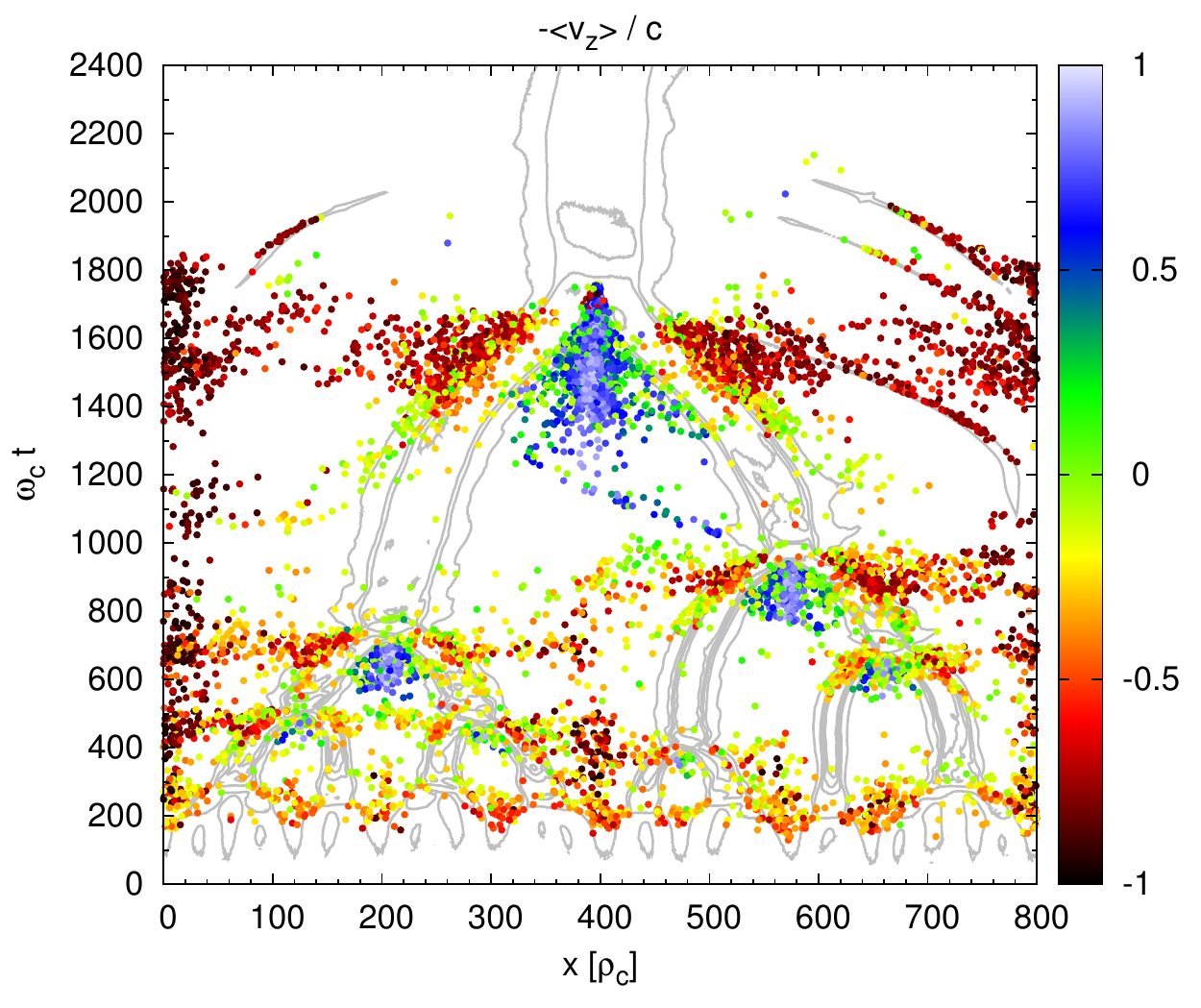}
\includegraphics[width=0.45\textwidth]{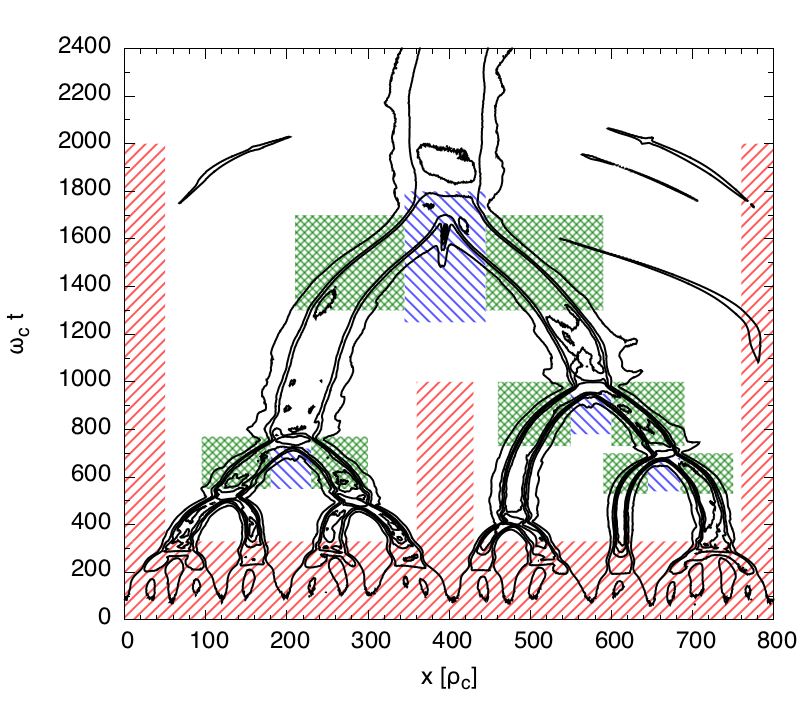}
\caption{Spacetime diagrams of the reconnection layer averaged over the region with $\Delta y = 20\rho_0$.\
\emph{Upper left} -- electron number density (arbitrary units).
\emph{Upper right} -- average electron Lorentz factor.
\emph{Middle left} -- $z$-component of the total electric field.
\emph{Middle right} -- $z$-component of the non-ideal electric field.
\emph{Lower left} -- particle locations at the beginning of the main acceleration episodes for a representative sample of electrons that exceed Lorentz factor of 20. The color indicates the mean velocity component $\left<v_z\right>/c$ during the main acceleration episode.
\emph{Lower right} -- Regions of the spacetime diagram used for classification of the main acceleration sites: magnetic X-points (\emph{blue}), plasmoid mergers (\emph{red}), and the plasmoids themselves (\emph{green}).
}
\label{fig_xtmap}
\end{figure*}

Figure~\ref{fig_xymap} suggests that the evolution of the reconnection layer can be projected onto the $x$ axis, preserving most of the useful information.
Therefore, it is convenient to represent the evolution of plasmoids on an $(x,t)$ spacetime diagram \citep[e.g.,][]{2009PhRvL.103j5004S,2011JGRA..116.9226F}.
To create a spacetime diagram, for a set of 500 regular time steps we integrate the parameters of interest within a stripe defined by $590\rho_0 \le y \le 610\rho_0$ or $\Delta y = 20\rho_0$, marked with white dotted lines on Figure~\ref{fig_xymap}.
At early stages of the simulation ($\omega_0t \lesssim 800 = L_x/\rho_0$) the integration along $y$ includes whole dense plasmoid cores, and at late stages of the simulation only the central portions of the plasmoids are included.

The upper left panel of Figure~\ref{fig_xtmap} shows the spacetime diagram $(x,t)$ of electron number density $n_{\rm e}$ (both drifting and background combined).
Many features that were identified in Figure~\ref{fig_xymap} --- the onset of the tearing mode instability, the hierarchical evolution of plasmoids, the emergence of secondary plasmoids, and even the formation of the secondary vertical current layer shortly before the final merger --- can be seen here at very high time resolution.
Some details of the plasmoid evolution are not represented on this diagram --- the shape of the plasmoids (elongated/circular), or the secondary plasmoids produced during the final merger and propagating along $-y$.
On the other hand, the spacetime diagram is very useful for describing the kinematics of plasmoids.
Because of the balanced evolution of plasmoids, each merger involves plasmoids of similar size and results in a cancellation of most $x$-momentum.
Correspondingly, the merged plasmoids are very slow, almost stationary.
Their bulk acceleration is not very rapid, and they only reach trans-relativistic velocities ($\Gamma\beta \sim 1$) shortly before a merger.
Only the small secondary plasmoids appear to be propagating with strongly relativistic velocities.
It is also apparent that the plasmoids attract each other with a force that is increasing with decreasing distance; this can be understood simply as due to Amp{\'e}re's force law.

The upper right panel of Figure~\ref{fig_xtmap} shows the spacetime diagram of the average electron Lorentz factor $\left<\gamma\right>$. This parameter combines the intrinsic energy distribution of electrons due to their random motions with their bulk motion. At $t = 0$ both the background and drifting electrons have Maxwellian distribution with $kT = m_{\rm e}c^2$, which corresponds to $\left<\gamma\right> \simeq 3.3$. In addition, however, the drifting electrons are boosted along the $z$-axis by drift velocity $\beta_{\rm d} = 0.6$, so the actual initial average Lorentz factor is $\left<\gamma\right> \simeq 4.3$.
For comparison, at the end of the simulation the average Lorentz factor of the background electrons is $\left<\gamma\right> \simeq 7.0$, hence the average background electron gained a Lorentz factor of $\left<\Delta\gamma\right> \simeq 3.6 = 0.22\sigma_0$.
As the plasmoids grow, they develop a distinct temperature structure (also noticed by \citealt{2014A&A...570A.111M}), being clearly hotter at the edges ($\left<\gamma\right> \simeq 12$) than in the center ($\left<\gamma\right> \simeq 8$).
A closer comparison with the density diagram reveals that high-temperature regions (\emph{plasmoid shells}) are located outside the high-density \emph{plasmoid cores}.
In the spacetime diagram, this structure is seen mainly for plasmoids with cores larger than $\Delta y$, in which case the diagram cuts across them.
Full $(x,y)$ maps of $\left<\gamma\right>$ (not shown) reveal that the high-temperature shells completely surround the high-density cores.
This structure is very persistent even after multiple plasmoid mergers, which indicates that the drifting particles that dominate the dense cores do not mix with the background particles energized in the magnetic X-points that dominate the hot shells.
The persistency of this structure could be specific to 2D simulations, where individual particles are constrained by conservation of the canonical momentum $P_z + (q/c) A_z$, where $P_z$ is the $z$-component of the particle momentum, and $A_z$ is the $z$-component of the magnetic vector potential, and hence they are confined to a given magnetic field line \citep[e.g.,][]{1998ApJ...509..238J}.
On the spacetime diagram of average Lorentz factor we can see additional weak structures induced after each merger and propagating far into the low-density regions.
Also in the low-density regions, we observe fan-like structures of particles being energized by the electric field in the magnetic X-points.
However, the average particle energies in the low-density regions are significantly lower than at the plasmoid edges.
High particle energies are also found in the secondary plasmoids propagating with relativistic velocities. They clearly stand out on the $\left<\gamma\right>$ diagram despite having low density contrast.

The middle left panel of Figure~\ref{fig_xtmap} shows the spacetime diagram of the $z$-component of the total electric field $E_z$.
We note that other components of the electric field -- $E_x,E_y$ -- take much smaller values along the reconnection midplane.
The middle right panel shows the $z$-component of the non-ideal electric field $(E_z + \left<\beta_x\right>B_y - \left<\beta_y\right>B_x)$, where $\left<\beta_i\right> = (\sum_{k=1}^N\beta_{i,k})/N$ is the mean value of the $i$-component of velocity of all $N$ particles (electrons and positrons) in each cell of interest.
Comparing these two panels, we note that the total electric field that determines particle acceleration is dominated by the ideal component (more specifically by $\left<\beta_x\right>B_y$) induced by the bulk motions of the plasmoids along the $x$-axis.
The non-ideal electric field forms three types of distinct structures:
 regions of negative reconnection field in the vicinity of the magnetic X-points,\footnote{The sign of the reconnection electric field $E_z$ in magnetic X-points is determined by the sign of the reconnecting magnetic field gradient ${\rm d}B_x/{\rm d}y$, which is opposite for the two layers in the simulation.} regions of positive anti-reconnection field between pairs of approaching plasmoids about to merge with each other, and rapidly propagating regions of positive field following each plasmoid merger on either side of the merged plasmoid.
As we will demonstrate in Section \ref{sec_res_accel}, these are some of the main sites of particle acceleration during our simulation, although many particles are also accelerated in plasmoids by ideal electric fields.

The lower panels of Figure~\ref{fig_xtmap} are discussed in Section \ref{sec_stat_sites}.

\subsection{Particle acceleration}
\label{sec_res_accel}

\subsubsection{Momentum distribution of all particles}

\begin{figure}[t]
\centering
\includegraphics[width=\columnwidth]{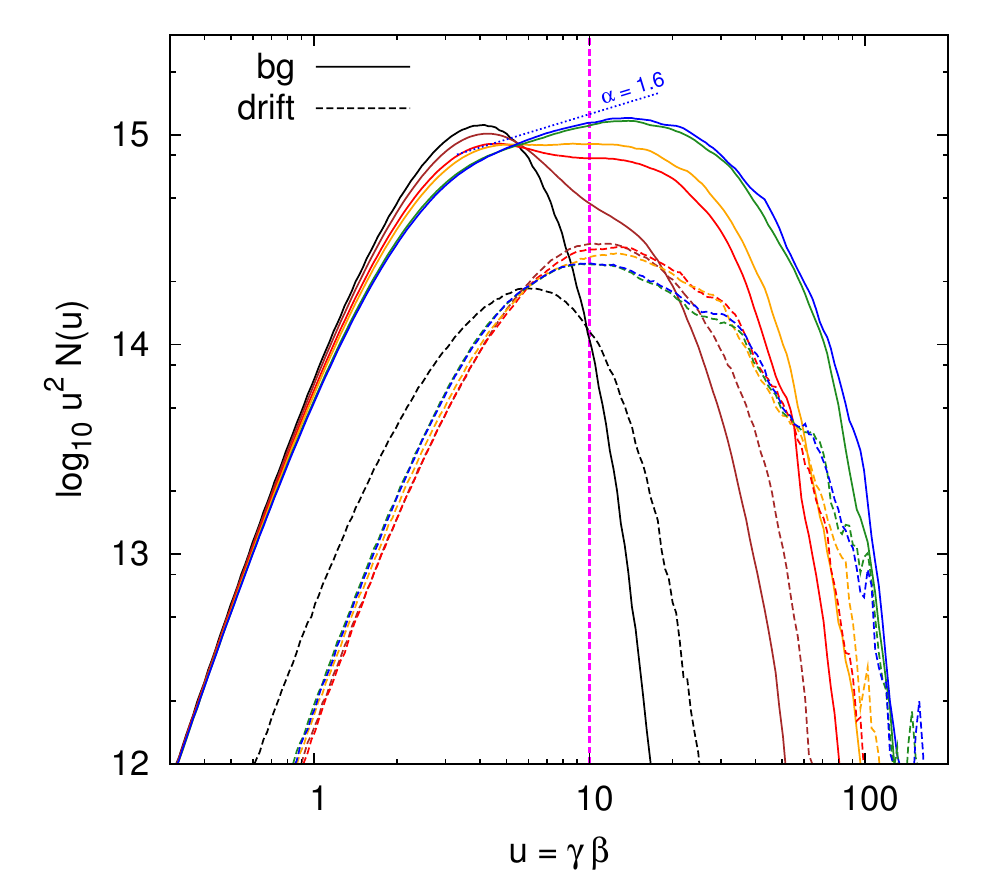}
\caption{Momentum distribution of all background (\emph{solid lines}) and drifting (\emph{dashed lines}) electrons for 6 simulation times: $t = 0$ (\emph{black}), $\omega_0t = 497$ (\emph{brown}), $\omega_0t = 994$ (\emph{red}), $\omega_0t = 1492$ (\emph{orange}), $\omega_0t = 1989$ (\emph{green}), and $\omega_0t = 2486$ (\emph{blue}). The energy limit $\gamma = 10$ for the selection of a representative sample of energetic background electrons is indicated with the \emph{vertical magenta line}.}
\label{fig_spectrum}
\end{figure}

In Figure~\ref{fig_spectrum} we show the time evolution of the spectrum $N(u)$ of background electrons, where $u = \gamma\beta$ is the dimensionless particle momentum (equivalent to particle energy for $u \gg 1$).
The spectrum is compensated by $u^2$ in order to reveal at which value of $u$ the particles contain most momentum.
The spectrum of the background positrons is identical to that of the background electrons.
The initial momentum distribution of background electrons is the Maxwell-J\"uttner distribution for dimensionless temperature $\Theta_{\rm b} = kT_{\rm b}/(m_{\rm e}c^2) = 1$. In time, the distribution extends into higher energies, developing an apparent power-law section of increasing hardness. Measuring the power-law index $\alpha$ defined by $N(u) \propto u^{-\alpha}$ is difficult because the power-law section is relatively short, and both the low-momentum and high-momentum cut-offs are relatively broad \citep[for more discussion see][]{2014arXiv1409.8262W}.
By considering a series of differently compensated spectra (not shown), we find that the final spectrum of background electrons ($\omega_0t = 2486$; solid blue line in Figure~\ref{fig_spectrum}) is consistent with $\alpha \simeq 1.6$, which is in agreement with the scaling results of \cite{2014arXiv1409.8262W} (see their Figure~3, noting that here we have $\sigma_0 = 16$ and $L/\sigma_0\rho_0 = 50$), and also with the results of \cite{2014ApJ...783L..21S} and \cite{2014PhRvL.113o5005G,2015ApJ...806..167G}.

\subsubsection{Main acceleration episode}
\label{sec_res_accel_epis}

Our main goal is to investigate the mechanisms of particle acceleration during the simulation by analyzing a representative sample of individual particles with their complete histories.
The total number of particles in our simulation is about $10^9$. 
It is unfeasible to record detailed history of each particle for analysis. Instead, we randomly selected a significant sample of $10^5$ \emph{tracked background electrons}, for which we recorded their detailed history for 500 time steps evenly distributed over the simulation time.
Of this sample, we are interested only in the highest-energy particles, and we arbitrarily choose a threshold minimum value of the Lorentz factor $\gamma_{\rm thr} = 10$.
As is illustrated in Figure~\ref{fig_spectrum}, this threshold is well above the peak of the initial momentum distribution of background electrons.
We selected a representative subsample of 14169 \emph{energetic particles} that exceeded $\gamma_{\rm thr}$ at least once during the simulation, and also at any time crossed the reconnection layer under consideration ($\tilde{y} = 0$, where $\tilde{y} = y - 0.75L_{\rm y}$). Using such a representative sample of energetic particles is essential to obtain a complete picture of acceleration mechanisms at work during the simulation.

For each energetic particle we identify what we call the \emph{main acceleration episodes}. They are defined as the shortest mutually exclusive intervals of time $[t_1:t_2]$ during which the particle gains energy $\gamma_i(t_2) - \gamma_i(t_1) \ge k(\gamma_{\rm i,max}-\gamma_{\rm i,min})$, where $\gamma_{\rm i,max},\gamma_{\rm i,min}$ are the global extrema of the particle energy history $\gamma_i(t)$, and $k$ is a constant.
Here we adopt $k = 0.5$ after experimenting with smaller values; hence during the main acceleration episode a particle gains energy that is at least one half of its total energy range during the simulation. A particle may have $m > 1$ non-overlapping main acceleration episodes since $\gamma(t)$ is usually non-monotonic. In such cases, each main acceleration episode will be treated as contributing to the particle maximum energy $\gamma_{\rm max}$ with $1/m$ weight. Each main acceleration episode is characterized by the initial location of the particle along the reconnection layer $x(t_1)$, the episode duration $\Delta t = t_2 - t_1$, and by the drift displacement along the $z$ axis $\Delta z = z(t_2) - z(t_1)$, or equivalently by the mean $z$ velocity $\left<v_z\right> = \Delta z / \Delta t$.

\begin{figure*}[ht]
\centering
\rotatebox{90}{\makebox[0.235\textwidth]{X-points}}
\includegraphics[width=0.235\textwidth]{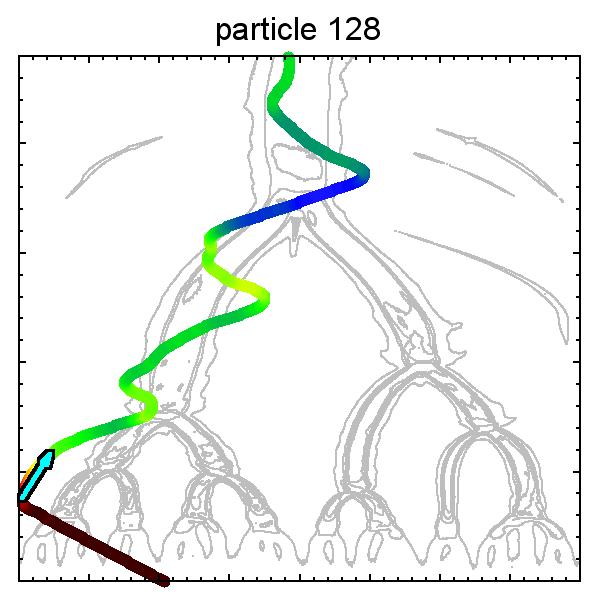}
\includegraphics[width=0.235\textwidth]{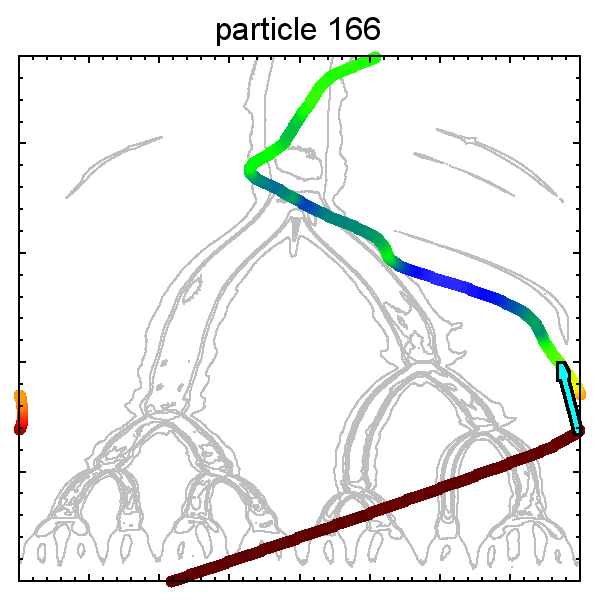}
\includegraphics[width=0.235\textwidth]{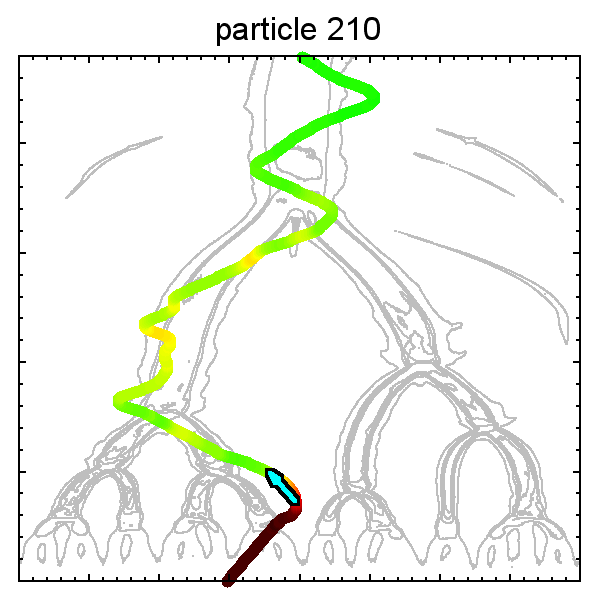}
\includegraphics[width=0.235\textwidth]{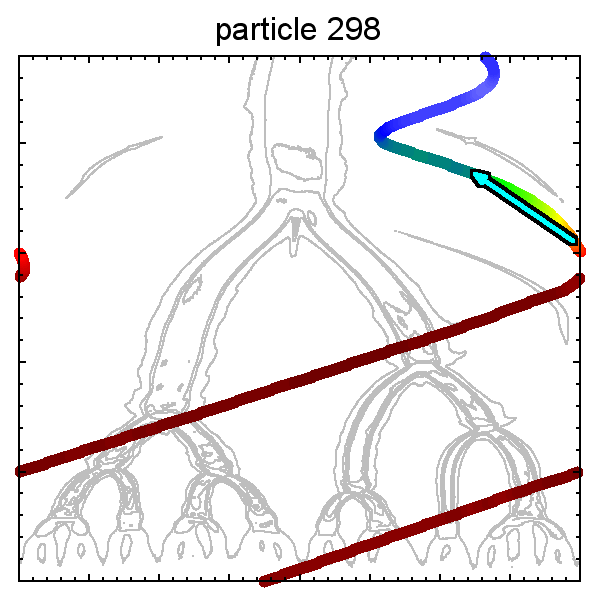}
\\
\rotatebox{90}{\makebox[0.235\textwidth]{mergers}}
\includegraphics[width=0.235\textwidth]{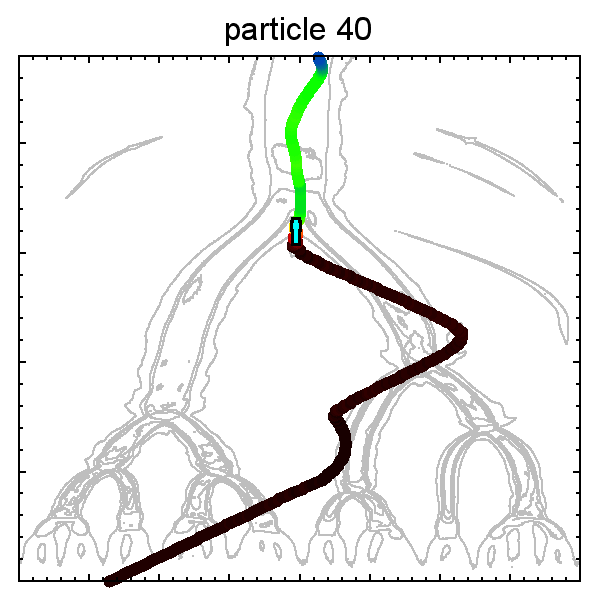}
\includegraphics[width=0.235\textwidth]{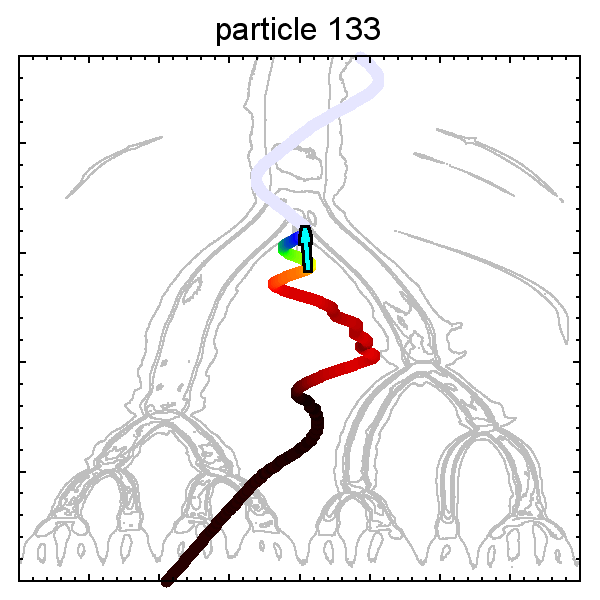}
\includegraphics[width=0.235\textwidth]{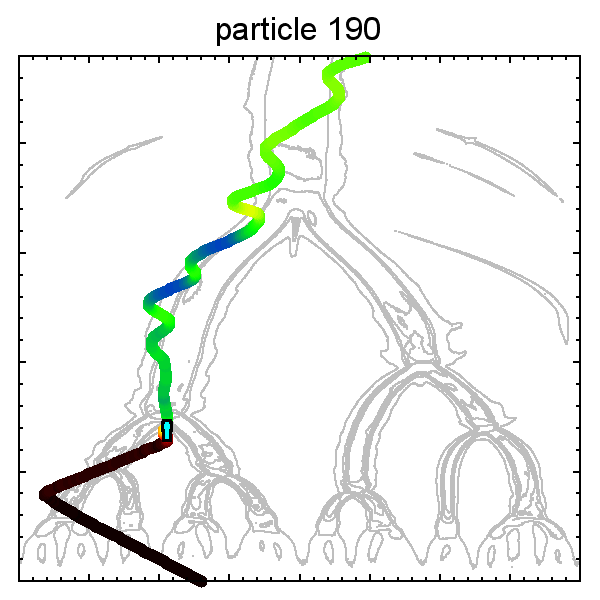}
\includegraphics[width=0.235\textwidth]{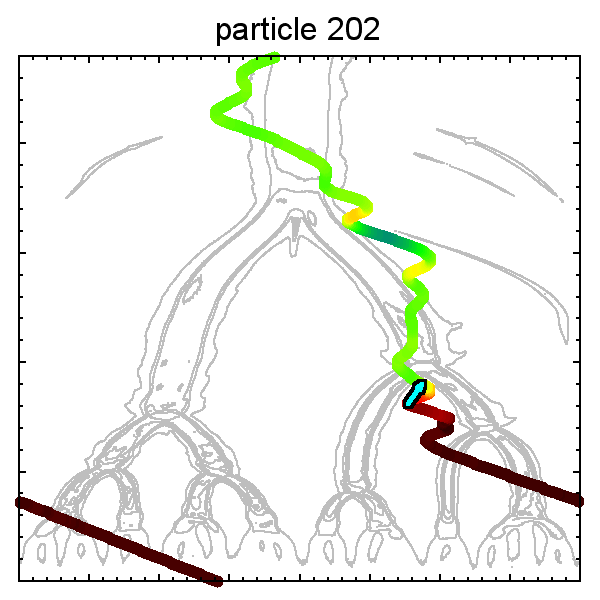}
\\
\rotatebox{90}{\makebox[0.235\textwidth]{plasmoids}}
\includegraphics[width=0.235\textwidth]{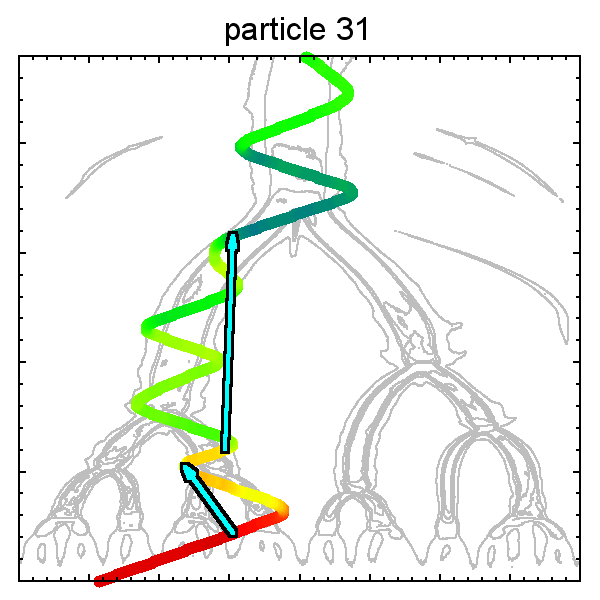}
\includegraphics[width=0.235\textwidth]{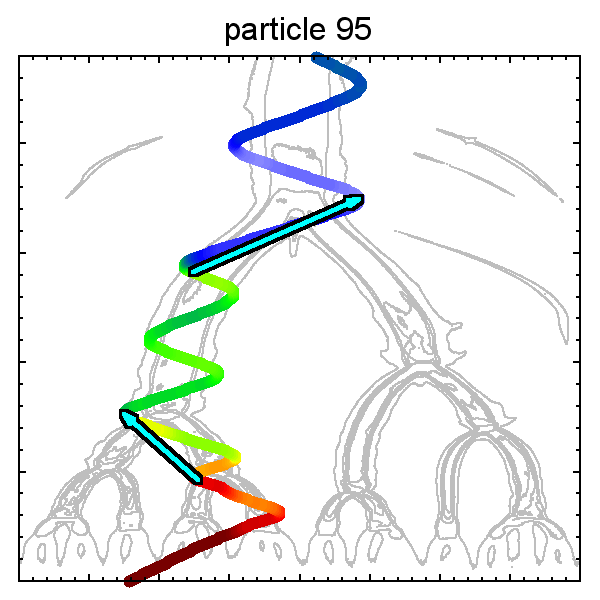}
\includegraphics[width=0.235\textwidth]{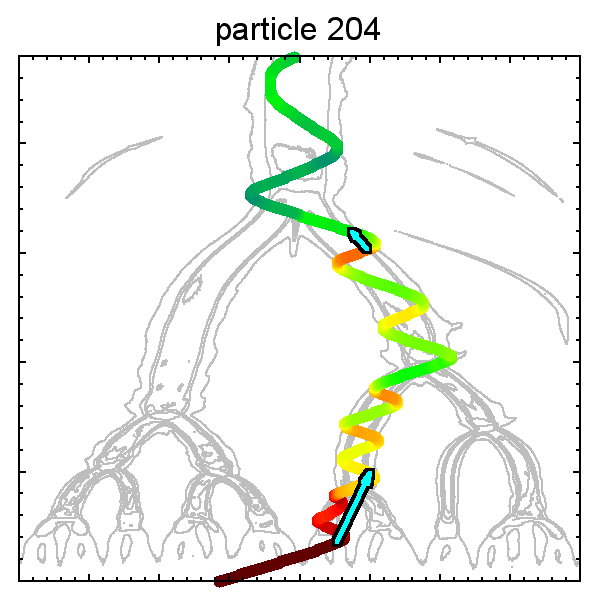}
\includegraphics[width=0.235\textwidth]{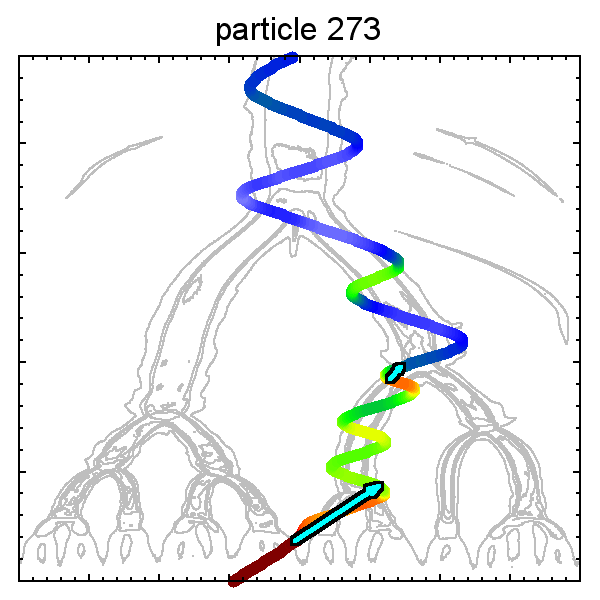}
\\
\rotatebox{90}{\makebox[0.235\textwidth]{other}}
\includegraphics[width=0.235\textwidth]{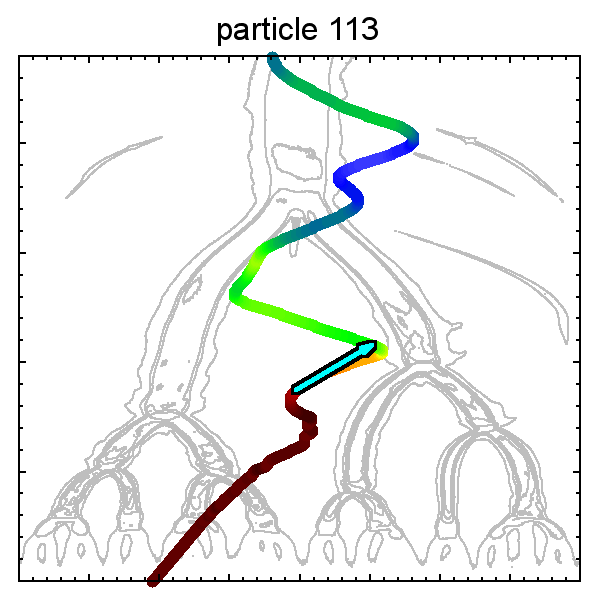}
\includegraphics[width=0.235\textwidth]{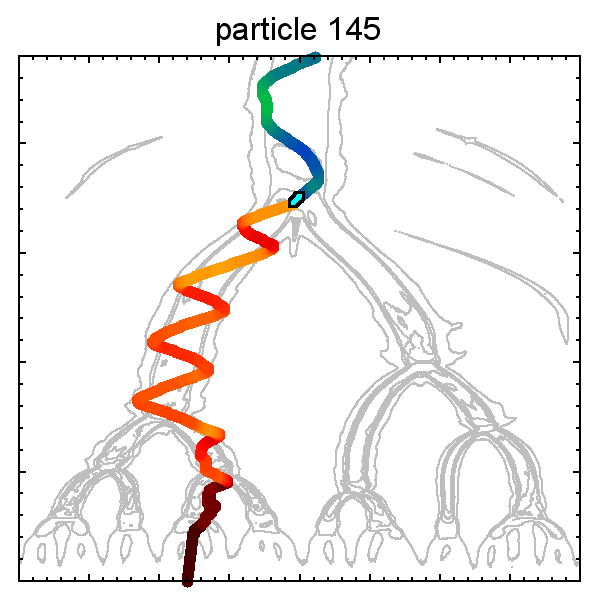}
\includegraphics[width=0.235\textwidth]{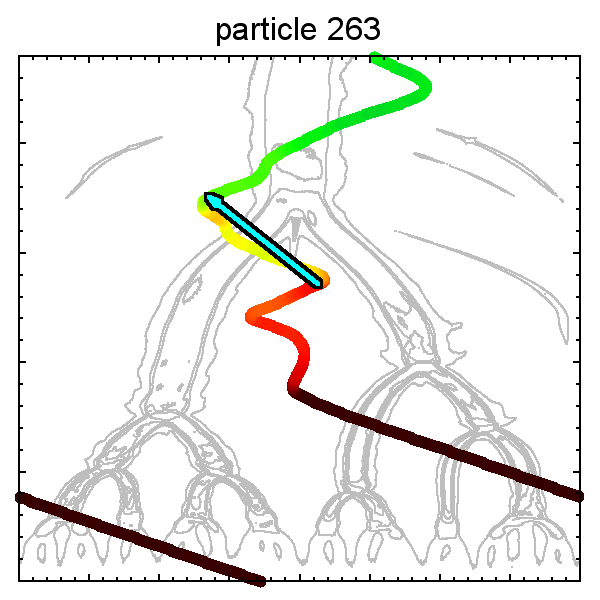}
\includegraphics[width=0.235\textwidth]{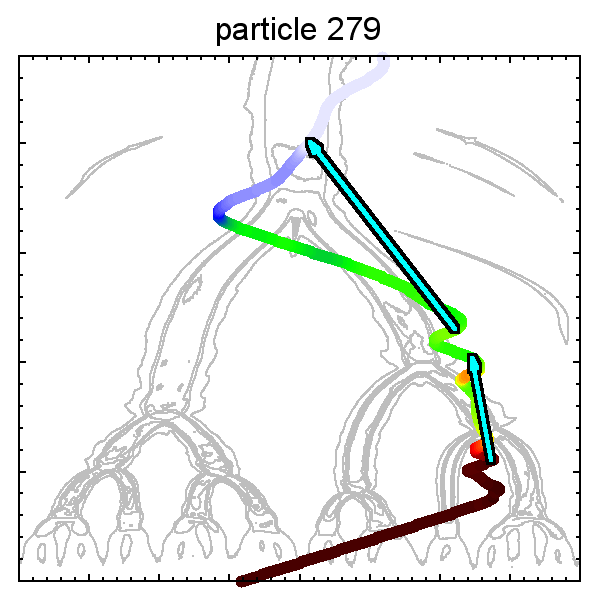}
\caption{Spacetime diagrams of the trajectories of selected individual electrons (color indicates instantaneous Lorentz factor; the color scale is the same as in Figure~\ref{fig_orbits_detail}), together with the number density contours (cf. the upper left panel of Figure~\ref{fig_xtmap}). \emph{Arrows} indicate the main acceleration episodes. The first row shows examples of particles accelerated by reconnection electric field at magnetic X-points. The second row shows examples of particles accelerated by anti-reconnection electric field between merging plasmoids. The third row shows examples of particles accelerated within the plasmoids. The fourth row shows other cases, including combinations of different mechanisms, and possibly acceleration at the post-merger disturbances.}
\label{fig_xtmap_orbits}
\end{figure*}

\subsubsection{Individual particle histories}

\begin{figure*}[ht]
\centering
\rotatebox{90}{\makebox[0.31\textwidth]{X-points}}
\includegraphics[width=0.69\textwidth]{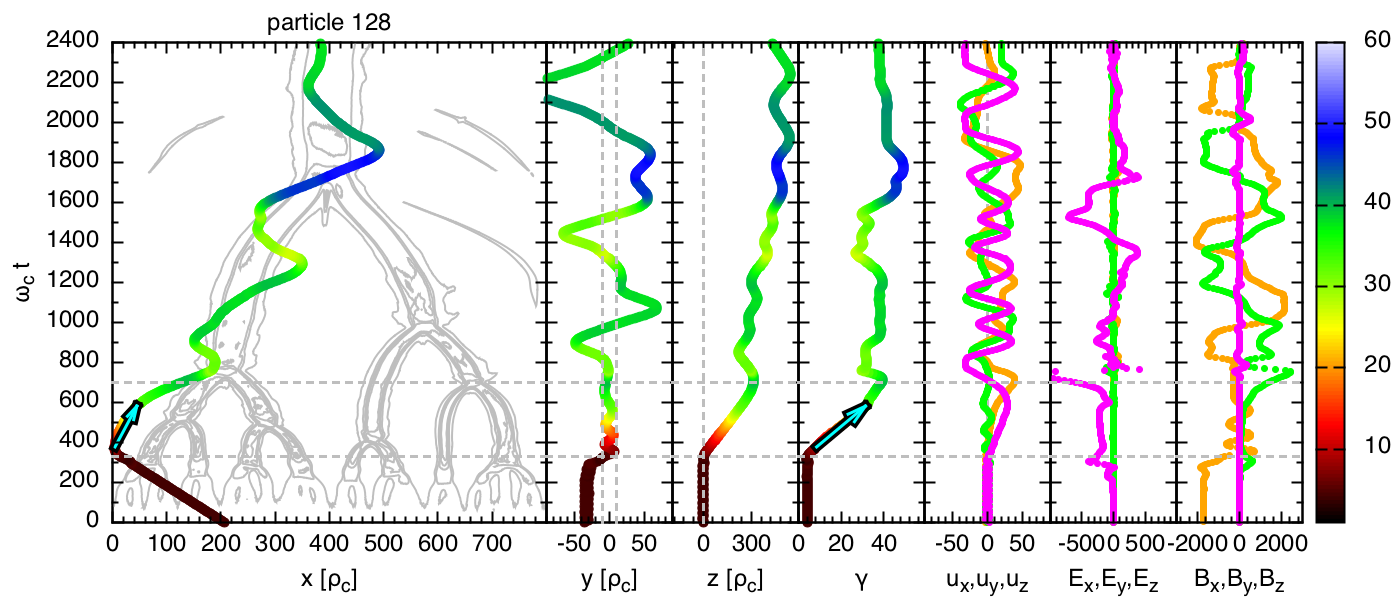}
\\
\rotatebox{90}{\makebox[0.31\textwidth]{mergers}}
\includegraphics[width=0.69\textwidth]{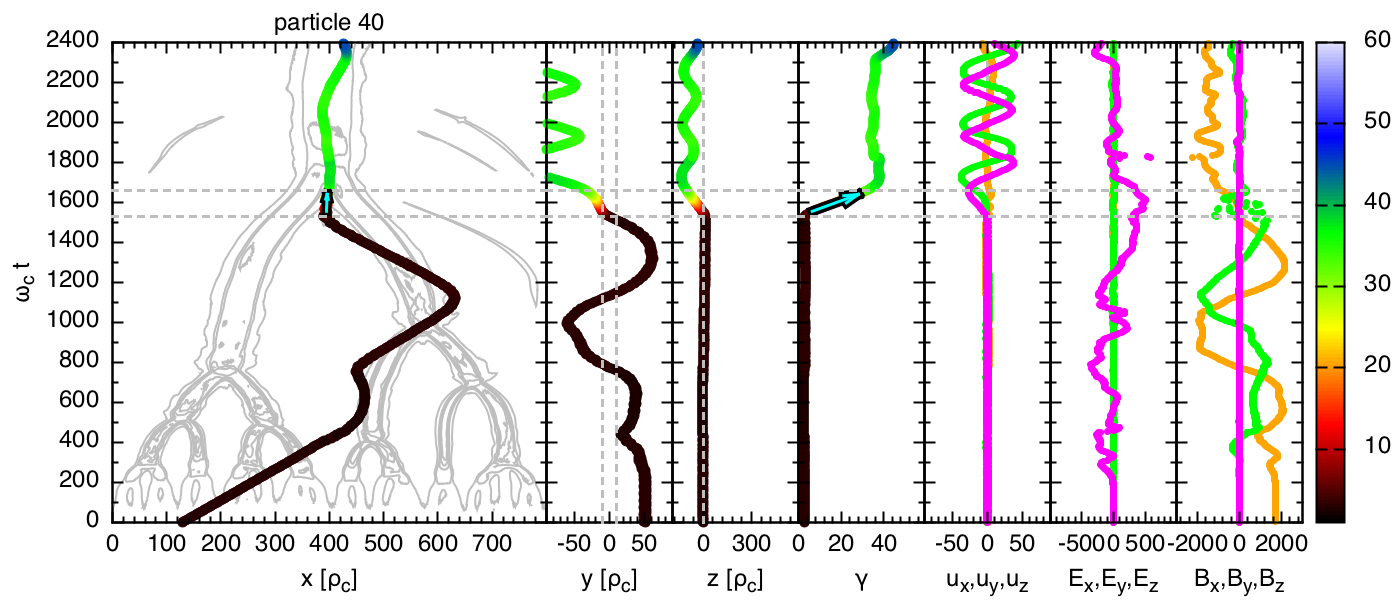}
\\
\rotatebox{90}{\makebox[0.31\textwidth]{plasmoids}}
\includegraphics[width=0.69\textwidth]{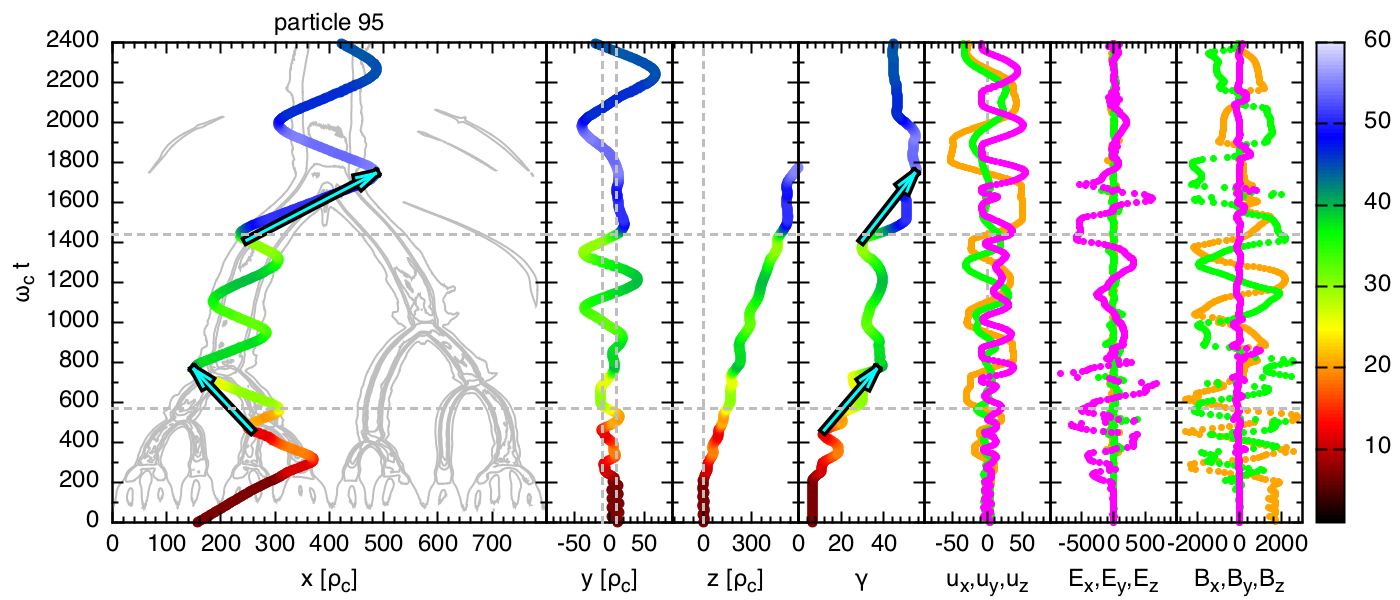}
\\
\rotatebox{90}{\makebox[0.31\textwidth]{other}}
\includegraphics[width=0.69\textwidth]{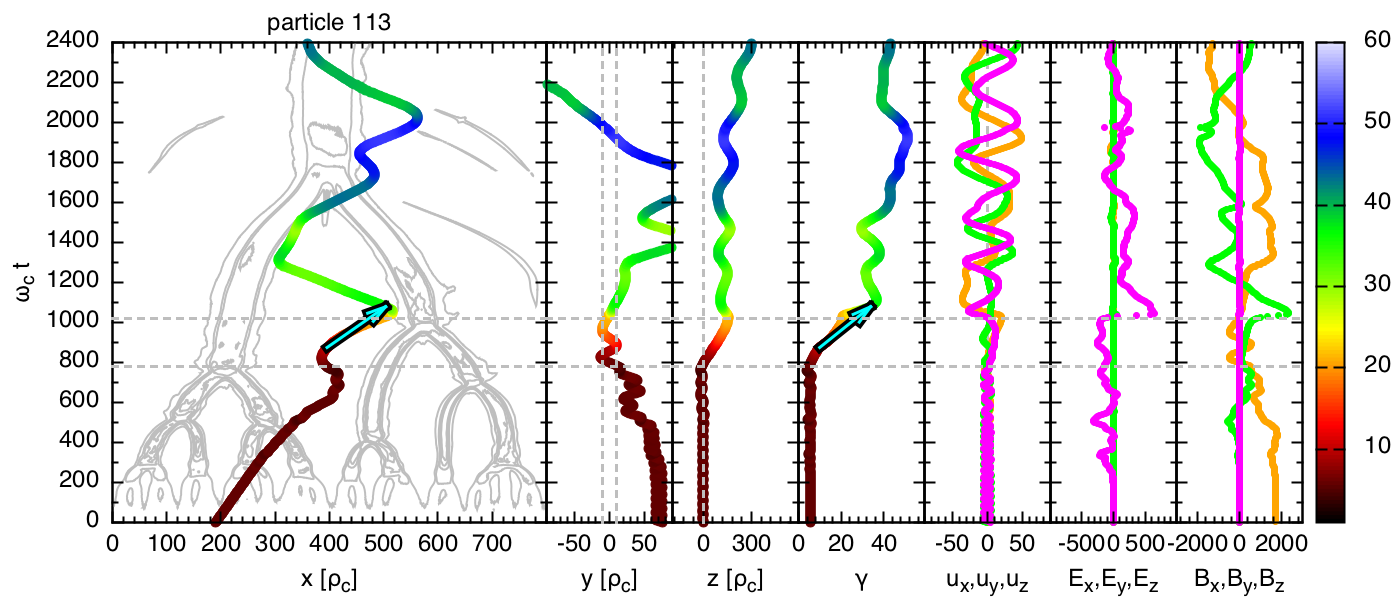}
\caption{Full 6-dimensional phase space and electromagnetic fields along the trajectories of selected individual electrons. The color scale used on the first 4 panels (from the left) and shown on the right indicates the electron Lorentz factor. On the last 3 panels, the $x$ components are \emph{orange}, $y$ components \emph{green}, and $z$ components \emph{magenta}. \emph{Arrows} indicate the main acceleration episodes. \emph{Horizontal dashed lines} indicate specific simulation times discussed for each particle in the main text.}
\label{fig_orbits_detail}
\end{figure*}

Before attempting to characterize this entire subsample, we will first consider selected individual examples. We can roughly classify the energetic particles into 4 groups, according to the location of their main acceleration site: (1) magnetic X-points, (2) regions between merging plasmoids, (3) plasmoids, (4) other/multiple sites. Each group is illustrated with 4 individual examples in Figure~\ref{fig_xtmap_orbits}, and one such example for each group is analyzed in detail in Figure~\ref{fig_orbits_detail}.

Particles belonging to group (1) gain most of their energy in the vicinity of magnetic X-points. A detailed analysis of a representative example (`particle 128') is shown in the first panel of Figure~\ref{fig_orbits_detail}.
The particle gains most energy from $\gamma \simeq 5$ at $\omega_0t = 330$ to $\gamma \simeq 39$ at $\omega_0t = 700$, starting at the major X-point at $x \simeq 0$ and drifting towards $x \simeq 100\rho_0$.
This corresponds well with the only main acceleration episode $\omega_0t \in [380:580]$.
At the same time, the particle is trapped in the reconnection layer ($\tilde{y} \in [-10:10]\rho_0$) and it drifts monotonically in the $+z$ direction on a Speiser-type orbit (\citealt{1965JGR....70.4219S}; note the oscillations in $\tilde{y}$ and $u_y$).
The particle feels roughly constant electric field $E_z \simeq -0.12B_0$ and oscillating magnetic field $B_x$ peaking at $\simeq \pm 0.3B_0$ (in phase with $\tilde{y}$).
One can roughly estimate that such electric field predicts a maximum energy gain of $\Delta\gamma \simeq |E_z/B_0|\omega_0\Delta t \simeq 44$, which is more than the actual energy gain $\Delta\gamma = 34$ of `particle 128'.
The particle $z$-momentum increases until $\omega_0t \simeq 600$, after which the particle starts interacting with the magnetic field $B_y$ (peaking at $\simeq 1.4B_0$ at $\omega_0t \simeq 750$) of the plasmoid centered at $x \simeq 140\rho_0$, which deflects it towards $+x$. At $\omega_0t \simeq 700$, the particle stops drifting towards $+z$ and the acceleration stops.
During $\omega_0t \in [330:700]$, the particle was displaced by $\Delta z \simeq 310\rho_0$, which means the average drift velocity $\left<\beta_z\right> = \Delta_z/(\omega_0\,\Delta t) \simeq 0.84$.
Subsequently, the particle follows the plasmoid on a wide slightly irregular orbit, and it gains additional energy, most likely through curvature drifts \citep[cf. Figure 6 in][]{2010ApJ...714..915O}.

Group (2) consists of particles that gain most of their energy between merging plasmoids.
A detailed analysis for a representative example (`particle 40') is presented in the second panel of Figure~\ref{fig_orbits_detail}.
The particle gains most energy from $\gamma \simeq 3$ at $\omega_0t = 1530$ to $\gamma \simeq 37$ at $\omega_0t = 1700$ between two merging plasmoids at $x \simeq 400\rho_0$.
This is captured by a main acceleration episode at $\omega_0t \in [1555:1640]$.
The particle passes through the reconnection layer midplane ($\tilde{y} \simeq 0$) exactly at $\omega_0t = 1530$, at which moment it starts drifting sharply towards the $-z$ axis.
The total displacement of the particle along $z$ is $\Delta z = -145\rho_0$, which means the average drift velocity of $\left<\beta_z\right> \simeq -0.85$.
The particle feels an anti-reconnection electric field varying in the range $E_z \sim (0.2-0.28)B_0$.
Such electric field predicts a maximum energy gain of $\Delta\gamma \simeq 41$, slightly higher than the actual value of $\Delta\gamma = 34$.
Consequently, the particle ends up located below the major plasmoid, oscillating in $\tilde{y}$ and $z$.

Group (3) consists of particles that gain most of their energy within plasmoids.
A detailed analysis for a representative example (`particle 95') is shown in the third panel of Figure~\ref{fig_orbits_detail}.
The particle gains energy gradually from $\gamma \simeq 7$ at $\omega_0t = 210$ to $\gamma \simeq 50$ at $\omega_0t = 1520$, for all this time being confined to plasmoids and following them in their evolution.
Two main acceleration episodes were identified at $\omega_0t \in [460:770]$ and $\omega_0t \in [1410:1750]$.
As the particle oscillates across the plasmoid, its energy fluctuates due to the superposition of the bulk motion of the plasmoid.
However, during some of the sharp turns in $u_x$ the particle gains more energy than it lost during the previous turn.
This is seen most clearly at $\omega_0t \simeq 570$ and $\omega_0t \simeq 1440$.
These two turns happen at the trailing edge of the plasmoid, where the particle experiences a strong electric field $E_z \sim -0.3B_0$ that is not balanced by the motional field since $u_x \simeq 0$ and $B_x \simeq 0$.
We tentatively identify this acceleration mechanism as the curvature drift \citep[e.g.,][]{2014PhPl...21i2304D,2015ApJ...806..167G,2015arXiv150502166L}.
In contrast to the particles accelerated in magnetic X-points (group 1) or between merging plasmoids (group 2), this particle does not experience a smooth drift along the $z$ axis. While there is a systematic drift towards $+z$ (with the average drift velocity $\left<\beta_z\right> \simeq 0.36$), the $z$-momentum ($u_z$) oscillates with a period twice shorter than oscillations of $u_x$ and $u_y$.
Shortly after the turns in $u_x$, most of the momentum gain is seen in the $x$ coordinate.
At the peaks of the $x$-momentum, the particle tends to cancel both its $y$- and $z$-momentum components.
The particle does briefly feel the anti-reconnection electric field $E_z \sim 0.3B_0$ at $\omega_0t \simeq 1620$, but the resulting additional acceleration is insignificant because of the very short time spent there.

Group (4) consists of particles for which the acceleration mechanism is different from the 3 discussed so far, or there is evidence for more than one acceleration mechanism during the particle history. The particular cases shown in Fig. \ref{fig_xtmap_orbits} include acceleration in the vicinity of plasmoids created in a recent merger (particle 113), acceleration between merging plasmoids for a particle previously bound in one of them (particle 145), and acceleration of unbound particles in the trailing edge of a plasmoid (particles 263 and 279), showing also that a particle accelerated between merging plasmoids can be accelerated further by a different mechanism.

In the fourth panel of Figure~\ref{fig_orbits_detail} we present a detailed analysis of `particle 113', accelerated after a nearby plasmoid merger but outside the merger product. The main acceleration episode is at $\omega_0t \in [880:1080]$. At $\omega_0t \simeq 780$, the particle enters a magnetic X-point at $x \simeq 400\rho_0$ with $\gamma \simeq 5$ and undergoes acceleration by the reconnection electric field $E_z \simeq -0.1B_0$, being displaced by $\Delta z \simeq 150\rho_0$.
At $\omega_0t \simeq 1030$ with $\gamma \simeq 21$, this mode of acceleration stops as the particle approaches a plasmoid newly formed in a recent merger.
The electric field switches sign and increases very sharply to $E_z \simeq 0.37B_0$, together with a strong increase in the magnetic field $B_y \simeq 1.35B_0$.
This forces the particle to bounce off the plasmoid and accelerate rapidly to $\gamma \simeq 36$ at $\omega_0t = 1100$.
The sharp increase in the electric field is different from that seen for particle 128 (representing acceleration at the magnetic X-points), which bounces off a plasmoid without additional acceleration. We observe additional structures in the electric field propagating outward after each plasmoid merger; these appear to be triggered by the merger, carrying excess momentum of the merging plasmoids, possibly via their Poynting flux (see the lower left panel of Figure~\ref{fig_xtmap}). We associate the sharp acceleration of particle 113 with its interaction with one of those disturbances.

\subsubsection{Statistics of the main acceleration sites}
\label{sec_stat_sites}

In order to evaluate the relative importance of the acceleration mechanisms discussed in the previous subsection, we have compared the parameters of all main acceleration episodes for our representative sample of energetic particles. The lower left panel of Figure~\ref{fig_xtmap} shows particle locations in the spacetime diagram corresponding to the beginning of each main acceleration episode. The color scale indicates the mean value of the $z$ velocity $\left<v_z\right>/c$ during the main acceleration episode, which is a measure of the dominant orientation of the electric field accelerating the particle. The spacetime distribution of particles undergoing significant acceleration is very complex. We observe several regions in the spacetime diagram where a large number of particles are accelerated. These regions are roughly and somewhat subjectively classified into three main types \emph{defined} on the lower right panel of Figure~\ref{fig_xtmap}: (1) particle diffusion regions around the magnetic X-points where $\left<v_z\right> > 0.5c$ (blue), (2) regions between merging plasmoids where $\left<v_z\right> < -0.5c$ (red), and (3) trailing edges of fast-propagating plasmoids where $\left<v_z\right> \gtrsim 0$ (green). Less important acceleration sites are associated with small secondary plasmoids in the later stages of the simulation ($\omega_0t > 1200$; blue), and also merger-triggered disturbances in the early stages of the simulation ($\omega_0t < 1200$).
In general, the acceleration of particles begins at $\omega_0t \simeq 200$, and is complete by $\omega_0t \simeq 2000$. And while it is widely spread in the spacetime diagram, it is not uniform. Very few particles are accelerated in the vicinity of static plasmoids, while the fast-propagating plasmoids accelerate many particles both at the leading and trailing edges (note the avoidance of dense plasmoid cores).

\begin{table*}
\caption{Parameters of function $N(\gamma) = N_0\gamma^{-\alpha}\exp[-(\gamma/\gamma_{\rm c})^q]$ fitted in the range $10 < \gamma < 150$ to the energy distributions of energetic particles according to their acceleration site (see Figure~\ref{fig_orbits_hist}). In order to calculate $\chi^2$, we add unity to the fitted function; the number of degrees of freedom is 12.}
\label{tab_fitparam1}
\vskip 1ex
\centering
\begin{tabular}{c|l|rrr|rrr}
\hline\hline
&& \multicolumn{3}{c|}{exponential ($q = 1$)} & \multicolumn{3}{c}{super-exponential ($q = 2$)} \\
k && $\alpha$ & $\gamma_{\rm c}$ & $\chi^2$ & $\alpha$ & $\gamma_{\rm c}$ & $\chi^2$ \\
\hline
    & X-points  & $-0.61 \pm 0.25$ &  $8.9 \pm 1.0$ & 14 & $0.61 \pm 0.15$ & $29.3 \pm 2.0$ & 31 \\
    & mergers   & $-0.05 \pm 0.12$ & $11.9 \pm 0.8$ & 10 & $0.89 \pm 0.08$ & $34.9 \pm 1.5$ & 53 \\
0.5 & plasmoids &  $0.40 \pm 0.18$ & $10.9 \pm 1.2$ & 15 & $1.31 \pm 0.08$ & $31.1 \pm 1.4$ & 11 \\
    & other     &  $1.57 \pm 0.15$ & $14.9 \pm 2.0$ & 16 & $2.21 \pm 0.08$ & $35.7 \pm 2.6$ & 11 \\
    & total     &  $0.64 \pm 0.12$ & $13.2 \pm 1.1$ & 33 & $1.43 \pm 0.06$ & $35.2 \pm 1.4$ & 85 \\
\hline
     & X-points  & $-0.66 \pm 0.23$ &  $9.0 \pm 0.9$ & 10 & $0.55 \pm 0.13$ & $29.4 \pm 1.7$ & 22 \\
     & mergers   &  $0.06 \pm 0.11$ & $12.8 \pm 0.8$ & 10 & $0.94 \pm 0.06$ & $36.4 \pm 1.3$ & 24 \\
0.25 & plasmoids &  $0.45 \pm 0.15$ & $11.9 \pm 1.1$ & 17 & $1.31 \pm 0.06$ & $33.2 \pm 1.4$ & 12 \\
     & other     &  $1.07 \pm 0.14$ & $12.3 \pm 1.2$ & 12 & $1.85 \pm 0.07$ & $32.5 \pm 1.9$ & 21 \\
     & total     &  $0.65 \pm 0.12$ & $13.2 \pm 1.1$ & 33 & $1.43 \pm 0.06$ & $35.3 \pm 1.4$ & 84 \\
\hline\hline
\end{tabular}
\end{table*}

\begin{figure}
\centering
\includegraphics[width=\columnwidth]{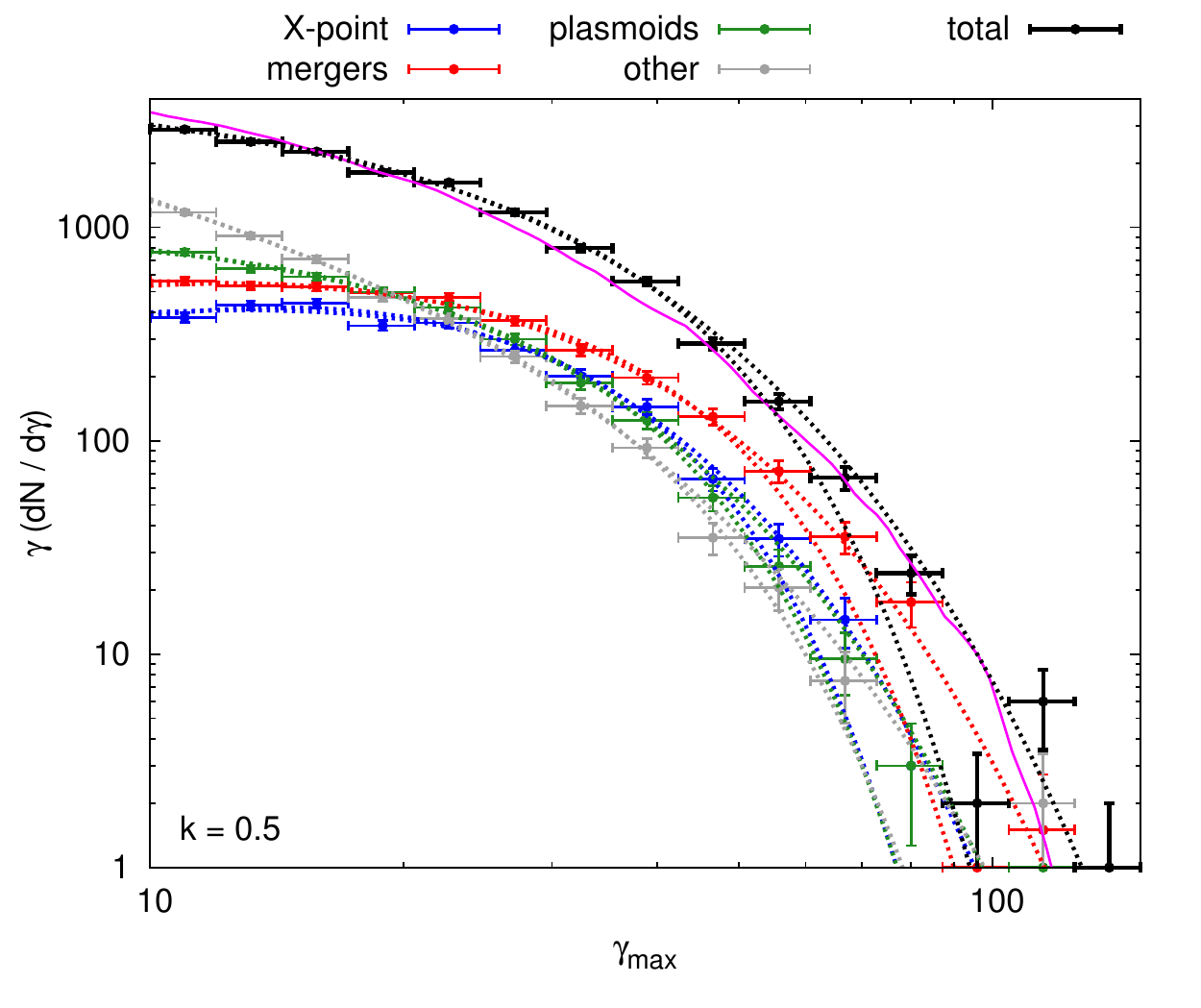}
\includegraphics[width=\columnwidth]{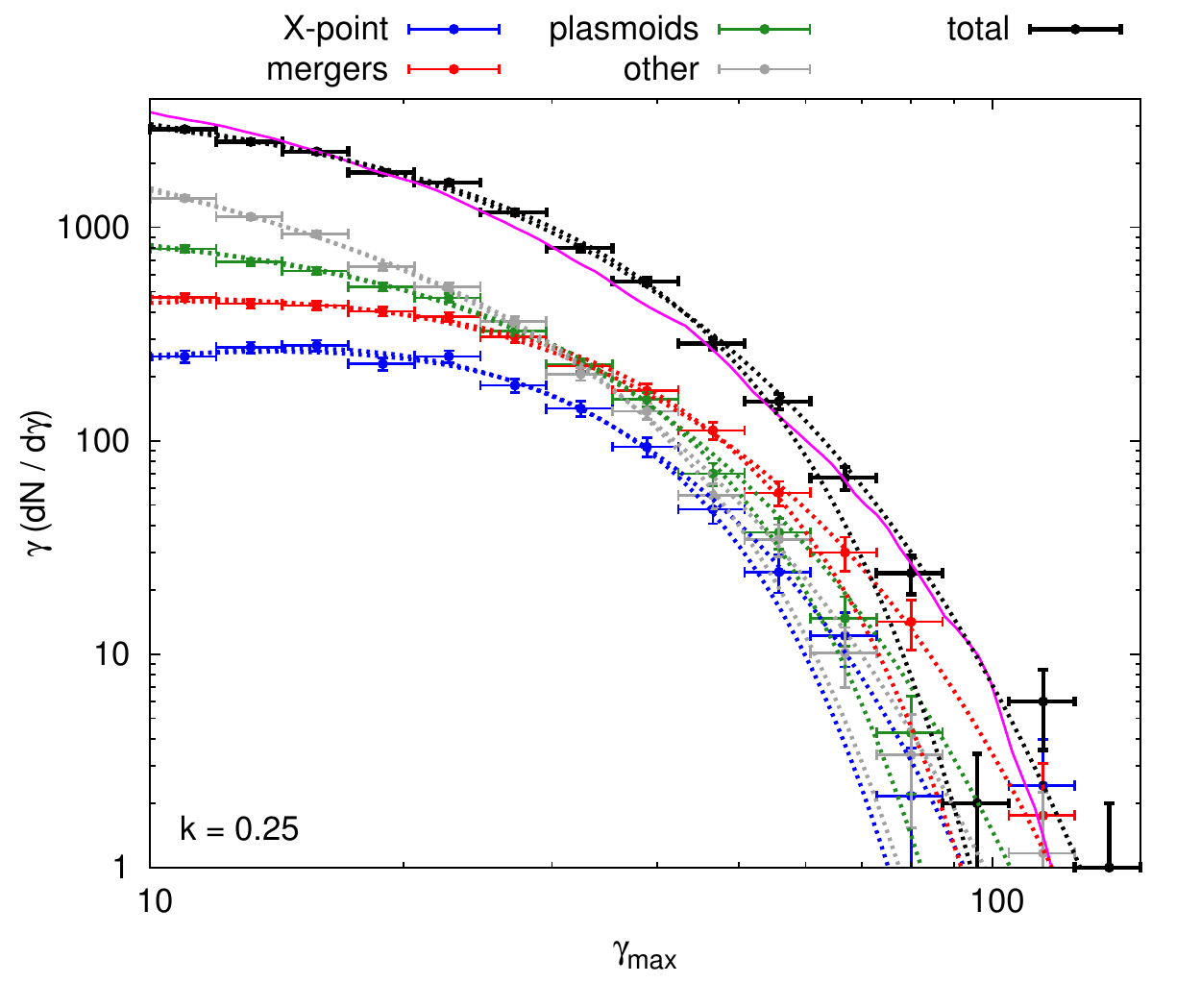}
\caption{Distribution of maximum particle energies for the sample of tracked energetic electrons (\emph{black}), divided according to our classification of main acceleration sites with $k = 0.5$ (top panel) and $k = 0.25$ (bottom panel): magnetic X-points (\emph{blue}), regions between merging plasmoids (\emph{red}), plasmoids (\emph{green}), and other sites (\emph{gray}). The vertical scale indicates the actual number of tracked electrons in each energy bin. Error bars are Poissonian only. Best-fit spectral models -- power-laws with exponential ($q = 1$) and super-exponential ($q = 2$) cut-offs (the high-energy tails of the $q = 1$ fits are always to the right from the corresponding $q = 2$ fits) -- are shown with dashed lines (see Table \ref{tab_fitparam1}). The solid magenta line shows a normalized momentum distribution of all background electrons at the end of the simulation.}
\label{fig_orbits_hist}
\end{figure}

In Figure~\ref{fig_orbits_hist}, we present the distribution of maximum particle energies for the representative sample of energetic particles in the energy range $10 < \gamma_{\rm max} < 150$. We also show separately the contributions from each class of the main acceleration sites, calculated for 2 values of parameter $k$. For each particle in the sample, we consider all $m$ main acceleration episodes, with each episode classified according to the initial location of the particle in the spacetime diagram (Figure~\ref{fig_xtmap}, bottom right panel). Each episode contributes with weight $1/m$ to the energy histogram at the maximum particle energy $\gamma_{\rm max}$. For a consistency check, we compare the histogram of $\gamma_{\rm max}$ values for the entire sample of energetic particles (gray points in Figure~\ref{fig_orbits_hist}) with the momentum distribution of all background electrons at the end of simulation (the magenta line). We find good agreement between the two distributions. The effect of changing parameter $k$ between 0.5 and 0.25 is modest and limited to $\gamma_{\rm max} < 40$.

For each class of the main acceleration sites, we performed a fit to the energy distribution of energetic electrons for $\gamma > 10$ with the following power-law model $N(\gamma) = N_0\gamma^{-\alpha}\exp[-(\gamma/\gamma_{\rm c})^q]$, with either exponential ($q = 1$) or super-exponential ($q = 2$) cut-off.
This approach is simpler than modeling using both types of cut-offs adopted in \cite{2014arXiv1409.8262W}, as our simulation is in the region of the $(L,\sigma_0)$ parameter space where neither cut-off shape is expected to dominate (see Section \ref{sec_dis}).
The best-fit parameters for $k = 0.5$ and $k = 0.25$ are listed in Table \ref{tab_fitparam1}, and all best-fit functions are shown in Figure~\ref{fig_orbits_hist}.
In the cases of mergers and other acceleration sites, the exponential cut-off provides a better fit, and in the cases of magnetic X-points and plasmoids the super-exponential cut-off is preferred. The values of the cut-off Lorentz factor are $\gamma_{\rm c} \sim 9-15$ for $q = 1$ and $\gamma_{\rm c} \sim 29-36$ for $q = 2$, regardless of the value of $k$.
Particles accelerated in magnetic X-points have the hardest distribution (lowest $\alpha$), followed by particles accelerated in mergers, within plasmoids and at other sites.
The value of $\alpha$ increases by $0.7-1.2$ between the fits with $q = 1$ and $q = 2$; since both types of cut-off can be acceptable, we cannot make a definite measurement of a power-law index of the electron energy distribution (cf. \citealt{2014arXiv1409.8262W}).
We find that in the energy range $10 < \gamma_{\rm max} < 20$, the largest contribution ($\sim 26\%$) to the total energy distribution is from the plasmoids, and in the energy range $40 < \gamma_{\rm max} < 100$, mergers clearly dominate (up to $\sim 50\%$) over other acceleration sites.
The contribution of magnetic X-points is no more than $\sim 25\%$, and the unclassified (`other') acceleration sites contribute between $\sim 10\%$ at $\gamma_{\rm max} \sim 70$ to $\sim 40\%$ at $\gamma_{\rm max} \sim 10$.

\begin{table}
\caption{Statistics of the number and classification of main acceleration episodes for different values of parameter $k$ and particle energy treshold $\gamma_{\rm min}$. For particles with 2 main acceleration episodes we report the most frequent sequences of the acceleration sites. Only values exceeding $3\%$ are reported.}
\label{tab_accelstat}
\vskip 1ex
\centering
\begin{tabular}{l|rr|r|r}
\hline\hline
k & \multicolumn{2}{c|}{0.5} & 0.33 & 0.25 \\
$\gamma_{\rm max}$ & $>10$ & $>30$ & $>30$ & $>30$ \\
\hline\hline
1 episode & 38\% & 57\% & $<3\%$ & 0\% \\
\hline
X-point   &  8\% & 14\% & --- & --- \\
merger    & 14\% & 29\% & --- & --- \\
plasmoid  &  3\% &  5\% & --- & --- \\
other     & 13\% & 10\% & --- & --- \\
\hline\hline
2 episodes              & 38\% & 38\% & 50\% & 7\% \\
\hline
X-point $\to$ merger    &  --- &  3\% &  4\% & --- \\
X-point $\to$ plasmoid  &  7\% & 12\% &  --- & --- \\
X-point $\to$ other     &  4\% &  5\% &  7\% & --- \\
merger $\to$ merger     &  3\% &  4\% & 14\% & 4\% \\
merger $\to$ plasmoid   &  --- &  4\% &  4\% & --- \\
plasmoid $\to$ plasmoid &  --- &  3\% &  --- & --- \\
other $\to$ merger      &  3\% &  --- &  4\% & --- \\
other $\to$ plasmoid    &  6\% &  4\% &  --- & --- \\
other $\to$ other       &  4\% &  --- &  4\% & --- \\
...                     &  ... &  ... &  ... & ... \\
\hline\hline
3 episodes    & 15\% &       5\%  & 33\% & 43\% \\
\hline\hline
$>3$ episodes &  8\% & $\sim 0\%$ & 14\% & 49\% \\
\hline\hline
\end{tabular}
\end{table}

In Table \ref{tab_accelstat}, we report the statistics of the number and classification of the main acceleration episodes for the representative sample of 14169 energetic particles with $\gamma_{\rm max} > 10$, and for a representative subsample of 1817 particles with $\gamma_{\rm max} > 30$ for three values of parameter $k = 0.5, 0.33, 0.25$.
In the case of $k = 0.5$, we find that the majority of energetic particles with $\gamma_{\rm max} > 10$ were accelerated either once or twice (38\% each).
The most important acceleration scenarios are: (1) single acceleration between merging plasmoids (14\%); (2) single acceleration in a magnetic X-point (14\%); (3) acceleration in a magnetic X-point followed by acceleration in a plasmoid (9\%).
The more energetic particles with $\gamma_{\rm max} > 30$ are more likely to be accelerated only once (57\%), particularly in a single merger event (29\%).
This is consistent with the relative contribution of mergers to the total energy distribution increasing with the particle maximum energy.
Two conclusions can be drawn from this: (1) while magnetic X-points and regions between merging plasmoids are the dominant acceleration sites, relatively few particles are accelerated in both of them; and (2) the particles that are accelerated within the plasmoids most likely were previously accelerated in a magnetic X-point.
This may help to explain why the contributions to the high-energy spectrum from magnetic X-points and plasmoids are comparable to each other, and different from the contribution from mergers.
For smaller values of $k$, almost all particles are characterized by multiple acceleration episodes. The most frequent sequence of two acceleration episodes is a sequence of two plasmoid mergers.

\section{Discussion}
\label{sec_dis}

Relativistic magnetic reconnection is a very complex phenomenon, where dissipation of magnetic energy, particle acceleration, and, under some conditions, high-energy emission proceed simultaneously. Investigating in detail the distribution and relative importance of the main acceleration sites is of the greatest importance for understanding the distribution of high-energy radiation emitted by these particles.
This radiation may be responsible for a variety of high-energy astrophysical phenomena including gamma-ray flares of blazars and the Crab Nebula, gamma-ray bursts, etc.
The characteristics of the radiation from plasmoid-dominated relativistic magnetic reconnection will be discussed in detail in a separate follow-up work.

By analyzing a representative sample of energetic particles, we have provided a comprehensive picture of particle acceleration for one simulation of relativistic reconnection. Other studies that have focused on investigating the histories of `representative' individual particles inevitably introduce subjective choices, which we have endeavored to avoid. While we do show representative examples for illustrative purposes (Figures \ref{fig_xtmap_orbits} and \ref{fig_orbits_detail}), we recognize that most individual particle histories are difficult to classify straightforwardly. For our main results, we rely on a simple concept of the main acceleration episodes, which can be easily identified for every particle. Other classification schemes could certainly be considered.

Our results indicate that in the plasmoid-dominated reconnection regime, there are three classes of main acceleration sites, together accounting for $\sim 80\%$ of the main acceleration episodes: magnetic X-points, regions between merging plasmoids (leading edges of fast-moving plasmoids), and trailing edges of fast-moving plasmoids.
That particles can be efficiently accelerated in relativistic magnetic X-points has been demonstrated previously in smaller simulations that did not capture multiple plasmoids \citep{2001ApJ...562L..63Z,2008ApJ...682.1436L}, as well as in previous large simulations \citep[e.g.,][]{2004PhPl...11.1151J,2005ApJ...618L.111Z,2007ApJ...670..702Z}.
That particles can be accelerated between merging plasmoids was demonstrated in the non-relativistic case \citep{2010ApJ...714..915O}, and mentioned briefly in the relativistic case by \cite{2014ApJ...783L..21S}.
\cite{2012ApJ...750..129B} argued that in addition to such acceleration, a significant number of particles are decelerated there.
We have not looked for evidence of such deceleration, however, we do not think that it would be in contradiction with our results.
As for the particles accelerated within plasmoids, they gain energy mostly when the plasmoids attain significant bulk motions, and do so mostly at the trailing edge.
This acceleration mechanism appears consistent with curvature drift \citep[e.g.,][]{2014PhPl...21i2304D,2015arXiv150502166L}, the importance of which was recently emphasized, in the context of relativistic reconnection, by \cite{2014PhRvL.113o5005G,2015ApJ...806..167G}.

By finding all main acceleration episodes for a representative sample of energetic particles, we are able to identify the most important acceleration scenarios.
According to Table \ref{tab_accelstat}, in the case of $k = 0.5$, the two most important acceleration sites are magnetic X-points and plasmoid mergers.
In the case of $k < 0.5$, most particles are accelerated in multiple acceleration episodes, and the most frequent sequence of acceleration sites is a sequence of 2 mergers.
We find that only $\sim 4\%$ of particles are characterized by a sequence of two main acceleration sites, the first an X-point, and the second a merger.
We have verified that while the fraction of particles characterized by this particular sequence increases with decreasing value of parameter $k$ in our definition of the main acceleration episode, in any case there are several other sequences that are more important than this one.
On the other hand, \cite{2014ApJ...783L..21S} found -- based on a much larger simulation -- that most of the highest-energy particles are energized in such sequences.
We suggest that the fraction of particles accelerated in plasmoid mergers that previously pass through a magnetic X-point may depend on the simulation size.

We do not find evidence of significant particle acceleration by a Fermi-type process operating in contracting plasmoids, as proposed by \cite{2006Natur.443..553D} in the case of non-relativistic reconnection in electron-ion plasma.
In that mechanism, an electron can be energized when circulating around the center of an elongated plasmoid by electric fields induced by a gradual contraction of the plasmoid.
In our simulation, the plasmoids formed in mergers are observed to relax on the dynamical time scale (i.e., Alfv\'en wave crossing time scale) and subsequently they do not undergo significant contraction.
On the other hand, the first generation of plasmoids forming directly from the tearing mode instability is characterized by significant elongation, as can be seen in Figure~\ref{fig_xymap} for $\omega_0t = 249$. These plasmoids circularize before $\omega_0t = 497$; however, few particles are significantly accelerated before this time (see the brown line in Figure~\ref{fig_spectrum}), and most of those are accelerated in magnetic X-points. 
We observe many energetic particles circulating within plasmoids --- a good example is `particle 95' shown in Figure
\ref{fig_orbits_detail}. However, these particles are significantly accelerated only when the plasmoids they are trapped in are themselves accelerated in bulk, as indicated by the spacetime distribution of the main acceleration episodes in the bottom right panel of Figure~\ref{fig_xtmap}.
We note that the evolution of plasmoids proceeds much faster in the relativistic case than in the non-relativistic case (since in the relativistic case, the Alfv\'en velocity is of the order of the speed of light).
Moreover, the initial episode of contracting islands is only a transient phenomenon, and hence this acceleration mechanism is less important in the case of relativistic reconnection relative to other mechanisms that depend on bulk plasmoid acceleration and mergers.
Plasma composition may also be important to this problem, as in pair plasma there is no separation of energy scales, and hence the Alfv\'en velocities, between different particle species.

Recent detailed studies of the total particle energy distribution resulting from relativistic reconnection are reported in \cite{2014ApJ...783L..21S,2014PhRvL.113o5005G,2014arXiv1409.8262W} and \cite{2015ApJ...806..167G}. The analysis of \cite{2014arXiv1409.8262W} suggests that the general form of the distribution is a power-law with an exponential [$\propto \exp(-\gamma/\gamma_{\rm c})$] and super-exponential [$\propto \exp(-(\gamma/\gamma_{\rm c})^2)$] cut-off, depending on the simulation domain size (an exponential cut-off is expected for simulation domains with $L_{x(y)} / \sigma_0\rho_0 \gtrsim 40$).
The power-law index $\alpha$ (${\rm d}N/{\rm d}\gamma \propto \gamma^{-\alpha}$) depends on the upstream magnetization $\sigma_0$ \citep[see also][]{2014ApJ...783L..21S,2014PhRvL.113o5005G,2015ApJ...806..167G}. For $\sigma_0 \gtrsim 10$, the distribution becomes very hard ($\alpha < 2$), with most energy contained around the high-energy cut-off which scales linearly with $\sigma$ or $L$.
In the case studied here --- $L_x = L_y = 800\rho_0$ and $\sigma_0 = 16$, hence $L_{x(y)} / \sigma_0\rho_0 = 50$ --- we expect to be at the boundary between the two regimes characterized by the predominance of the exponential or super-exponential cut-offs.
Indeed, we find that both types of cut-off can describe contributions to the energy distributions of energetic particles from different types of main acceleration sites (see Figure~\ref{fig_orbits_hist} and Table \ref{tab_fitparam1}).
We also find that the compensated momentum distribution $u^2N(u)$ of all background electrons peaks between $10 < u_{\rm peak} < 20$ at late stages of the simulation, with the effective power-law index $\alpha \simeq 1.6$ (see Figure~\ref{fig_spectrum}).

In principle, the energy distribution of particles could extend to even higher energies in a steeper power-law tail if at least some particles could undergo multiple energizations reminiscent of a first-order Fermi process \citep{2012PhRvL.108m5003H}. Our analysis reveals that about 10\% of energetic particles experience more than 2 main acceleration episodes. However, due to our definition of an acceleration episode ($\Delta\gamma \sim k\gamma_{\rm max}$ with $k = 0.5$), those particles must undergo significant deceleration between every two accelerations, and the history of those particles is not consistent with the canonical Fermi process ($\Delta\gamma \sim \gamma$).

Previous simulations performed -- unlike here -- in the radiatively efficient regime, where the synchrotron cooling time scale for the most energetic particles is comparable to or shorter than the dynamical time scale, emphasized particle acceleration in magnetic X-points on Speiser-type orbits as the most important acceleration mechanism \citep{2013ApJ...770..147C,2014ApJ...782..104C}.
In the radiatively efficient regime, particle acceleration is limited in the presence of a strong perpendicular magnetic field.
These works demonstrated that particles accelerated in magnetic X-points experience only weak perpendicular magnetic field components.
On the other hand, one can expect stronger perpendicular magnetic field components in the plasmoids.
Hence, in the radiatively efficient regime acceleration within plasmoids or in plasmoid mergers may be suppressed by the severe radiative losses.
This issue requires a further, more detailed investigation.

Finally, the simulation presented in this work uses no initial magnetic perturbation and periodic boundary conditions, which leads to a rapid hierarchical evolution of the plasmoids via balanced mergers that tend to cancel the bulk of the $x$-component of momentum.
The distribution of the particle acceleration sites may be fundamentally different with open boundary conditions. 
Such simulations tend to develop a roughly statistically steady-state configuration (with some tearing) with a central magnetic X-point and broad Alfv\'enic outflows \citep{2007PhPl...14g2303D,2012PhPl...19d2303L}.
We would then expect a significantly smaller role of plasmoid mergers in accelerating particles and producing sharp radiation flares.

\section{Conclusions}
\label{sec_con}

We presented a study of the evolution of a plasmoid-dominated reconnection layer and of the distribution and relative importance of the main particle acceleration sites by analyzing a kinetic numerical simulation of relativistic magnetic reconnection in pair plasma with initial magnetization $\sigma_0 = 16$ and no guide field.
Substantial insight is gained by analyzing spacetime diagrams for various physical parameters, and by investigating a representative sample of tracked energetic particles (see Figure~\ref{fig_xtmap}).
We find that the plasmoids develop a robust intrinsic structure with colder dense cores and hotter outer shells that is preserved through multiple plasmoid mergers.
After the initial episode of elongated plasmoids arising from the tearing mode instability, the plasmoids are not contracting noticeably.
We identify three major types of particle acceleration sites: (1) magnetic X-points with the reconnection electric field $E_z < 0$ (the sign refers to this particular case), (2) regions between merging plasmoids with the anti-reconnection electric field $E_z > 0$, and (3) the trailing edges of accelerating plasmoids.
Notably, particle acceleration is not observed around static plasmoids, and it cannot be associated with the initial transient phase of plasmoid contraction.
We introduce the concept of the main acceleration episode as the shortest period of time $[t_1:t_2]$ where $\gamma(t_2)-\gamma(t_1) > k(\gamma_{\rm max}-\gamma_{\rm min})$, and use it to compare the relative importance of the different classes of acceleration site in producing the high-energy electron distribution.
For $k = 0.5$, we find that plasmoid mergers are able to accelerate particles to higher energies ($\gamma \sim 100$), with a harder distribution, than magnetic X-points or accelerating plasmoids (the power-law index differs by $\sim 0.35$, but its actual value depends on the assumed shape of the high-energy cut-off; see Figure~\ref{fig_orbits_hist} and Table \ref{tab_fitparam1}).
As a consequence, particles accelerated in plasmoid mergers dominate the energy distribution for $\gamma > 40$.
The particles accelerated during plasmoid mergers are usually not the same particles that are accelerated at other sites (see Table \ref{tab_accelstat}).
The particles accelerated in accelerating plasmoids were usually also accelerated in magnetic X-points.
These results will be essential for interpreting the time-, angle- and energy-dependent radiative signatures of relativistic magnetic reconnection, which will be described in a future study.

\acknowledgments 

This work was supported in part by NASA Astrophysics Theory Program grant NNX14AB37G, NSF grant AST-1411879, DoE grant DE-SC0008409, and NASA Fermi Guest Investigator Program.
K.N. was supported by NASA through Einstein Postdoctoral Fellowship grant number PF3-140130 awarded by the Chandra X-ray Center, which is operated by the Smithsonian Astrophysical Observatory for NASA under contract NAS8-03060.
B.C. acknowledges support from the Lyman Spitzer Jr. Fellowship awarded by the Department of Astrophysical Sciences at Princeton University, and the Max-Planck/Princeton Center for
Plasma Physics.
We acknowledge the use of computational resources obtained from XSEDE (Stampede), University of Colorado Research Computing (Janus), and the local cluster Verus at the Center for Integrated Plasma Studies at the University of Colorado Boulder.

\end{document}